\documentclass[letterpaper,twocolumn,twoside]{IEEEtran}
\usepackage{amssymb, cite, epsfig}

\def\beq{\begin{equation}}
\def\eeq{\end{equation}}
\def\beqa{\begin{eqnarray}}
\def\eeqa{\end{eqnarray}}
\def\beqan{\begin{eqnarray*}}
\def\eeqan{\end{eqnarray*}}
\def\C{{\mathbb{C}}}
\def\argmin{\mathop{\mathrm{arg\,min}}}
\def\dim{\mathop{\mathrm{dim}}}
\def\range{\mathop{\mathrm{range}}}

\def\limn{\lim_{n \rightarrow \infty}}
\def\liminfn{\liminf_{n \rightarrow \infty}}
\def\limsupn{\limsup_{n \rightarrow \infty}}

\newtheorem{definition}{Definition}
\newtheorem{theorem}{Theorem}
\newtheorem{lemma}{Lemma}
\newtheorem{algorithm}{Algorithm}
\def\Ihat{\ensuremath{\hat{I}}}
\def\IhatSOMP{\ensuremath{\hat{I}_{\rm SOMP}}}
\def\Itrue{\ensuremath{I_{\rm true}}}
\def\Perr{\ensuremath{p_{\rm err}}}
\def\PMD{\ensuremath{p_{\rm MD}}}
\def\PFA{\ensuremath{p_{\rm FA}}}
\def\gammaConst{\ensuremath{\gamma_{\rm const}}}
\def\gammaOpt{\ensuremath{\gamma_{\rm opt}}}
\def\SNR{\mbox{\small \sffamily SNR}}
\def\MAR{\mbox{\small \sffamily MAR}}
\def\tableSNR{\mbox{\tiny \sffamily SNR}}
\def\tableMAR{\mbox{\tiny \sffamily MAR}}
\def\captionSNR{\mbox{\scriptsize \sffamily SNR}}
\def\captionMAR{\mbox{\scriptsize \sffamily MAR}}
\def\arr{\rightarrow}
\def\Exp{\mathbf{E}}
\def\sigmahat{\ensuremath{\widehat{\sigma}}}
\def\SNRmin{\mbox{\small \sffamily SNR}_{\rm min}}
\def\dB{{\rm dB}}
\def\BetaDist{\mbox{Beta}}
\def\Ptrue{\ensuremath{\mathbf{P}_{\rm true}}}
\def\rhotrue{\ensuremath{\rho_{\rm true}}}
\def\mmin{m_{\rm min}}
\def\smin{s_{\rm min}}
\def\zmax{s_{\rm max}}

\newcommand{\abf}{\mathbf{a}}
\newcommand{\ebf}{\mathbf{e}}

\newcommand{\vbf}{\mathbf{v}}
\newcommand{\wbf}{\mathbf{w}}
\newcommand{\xbf}{\mathbf{x}}
\newcommand{\xbfhat}{\widehat{\mathbf{x}}}
\newcommand{\ybf}{\mathbf{y}}
\newcommand{\Abf}{\mathbf{A}}
\newcommand{\Pbf}{\mathbf{P}}

\newcommand{\etal}{\emph{et al.}}

\begin{document}

\title{On--Off Random Access Channels: \\ A Compressed Sensing Framework}
\author{Alyson K. Fletcher, 
        Sundeep Rangan,
        and~Vivek~K~Goyal 
\thanks{This work was submitted in part to the IEEE Int.\ Symp.\ on
        Information Theory, Seoul, Korea, June--July 2009.}%
\thanks{A. K. Fletcher (email: alyson@eecs.berkeley.edu) is a
        postdoctoral researcher with Prof.\ Martin Vetterli at
        the Department of Electrical Engineering and Computer Sciences,
        University of California, Berkeley.}
\thanks{S. Rangan (email: srangan@qualcomm.com) is with Qualcomm Technologies,
        Bedminster, NJ.}%
\thanks{V. K. Goyal (email: vgoyal@mit.edu) is with
        the Department of Electrical Engineering and Computer Science and
        the Research Laboratory of Electronics,
        Massachusetts Institute of Technology.
        His work was supported in part by
        NSF CAREER Grant CCF-643836\@.}}

\markboth{On--Off Random Access Channels: A Compressed Sensing Framework}
        {Fletcher, Rangan and Goyal}

\maketitle

\begin{abstract}
This paper considers a simple on--off random multiple access channel,
where $n$ users communicate simultaneously to a single receiver over
$m$ degrees of freedom.
Each user transmits with probability $\lambda$,
where typically $\lambda n < m \ll n$,
and the receiver must detect which users transmitted.
We show that when the codebook has i.i.d.\ Gaussian entries,
detecting which users transmitted is
mathematically equivalent to a certain sparsity detection problem
considered in compressed sensing.
Using recent sparsity results,
we derive upper and lower bounds on the capacities of these channels.
We show that common sparsity detection algorithms,
such as lasso and orthogonal matching pursuit (OMP),
can be used as tractable multiuser detection schemes
and have significantly better performance than single-user
detection.
These methods do achieve some near--far resistance but---at high
signal-to-noise ratios (SNRs)---may achieve capacities
far below optimal maximum likelihood detection.
We then present a new algorithm, called sequential OMP, that illustrates
that iterative detection combined with power ordering or power shaping can
significantly improve the high SNR performance.
Sequential OMP is analogous to successive interference cancellation
in the classic multiple access channel.
Our results thereby provide insight into the roles of
power control and multiuser detection on random-access signalling.
\end{abstract}


\section{Introduction}
In wireless systems, \emph{random access} refers to any
multiple access communication protocol
where the users autonomously decide whether or not to transmit
depending on their own
traffic requirements and estimates of the network load.
While random access is best known for its use
in packet data communication in
wireless local area networks (LANs) \cite{CallawayGHGNHB:02},
this paper considers random access for simple on--off messaging.
On-off random access signaling can be used for a variety of control
tasks in wireless networks such as user
presence indication, initial access, scheduling requests and paging.
Random on--off signaling is already used for some of these
tasks in current cellular systems
\cite{BenderBGPSV:00,HolmaT:06}

The limits of on--off random access signaling with multiple users
are not fully understood.
To this end, we consider a simple
random multiple access channel where $n$ users transmit to
a single receiver.  Each user is assigned a single codeword which it
transmits with probability $\lambda$.
We wish to understand the capacity of these channels,
by which we mean the total number of degrees of freedom
$m$ needed to reliably detect which users transmit
as a function of $n$, $\lambda$, and the channel conditions.
We also wish to establish performance bounds for specific
decoding algorithms.

This on--off random access channel is related to the classic
multiple access channel (MAC) in network information theory
\cite{Ahlswede:71,CoverT:91}.
The theory of the MAC channel is well understood
\cite{Verdu:86,VerduS:99,TseH:99,HonigMV:95} and has been applied in
commercial CDMA systems \cite{Andrews:05}.
Unfortunately, it is difficult to
apply the classic MAC channel analysis
directly to the on--off random access channel under consideration here.

In the traditional analysis of the MAC channel,
the number of users remains constant,
while the number of degrees of freedom of the channel goes to infinity.
As a result, each user can employ a capacity-achieving code with an infinite
block length.
However, in the on--off random access channel considered here,
as the number of degrees of freedom of the channel is increased, the goal is not
to scale the number of bits per user, but rather the total number of users.
Since each user only transmits at most one bit of information,
channel coding cannot be used for reliability,
and the classic MAC capacity results do not apply.

Our analysis is instead based on identifying a connection between
the on--off random access channel and the recovery of the sparsity pattern of a signal from noisy random linear measurements.
The feasibility of recovering sparse, approximately sparse,
or compressible signals from a relatively small number of
random linear measurements has recently been termed
\emph{compressed sensing} \cite{CandesRT:06-IT,Donoho:06,CandesT:06}.
When the users in the on--off random access channel
employ certain large random codebooks,
we show that the problem at the receiver of detecting
the active users is precisely the sparsity detection
problem addressed in several recent works
in the compressed sensing literature
\cite{Wainwright:07-725,FletcherRG:08arXiv,Wainwright:06,TroppG:07}.

Results in compressed sensing generally provide
bounds on the $\ell^2$ estimation error of a signal
as a function of the number of measurements, the
signal sparsity and other factors.
However, what is relevant for the random on--off multiple access
channel is detecting the \emph{positions} of the nonzero entries.
This problem arises in subset selection in linear regression~\cite{Miller:02}.

By exploiting recent compressed sensing
results and providing an analysis of a new algorithm,
we are able to provide a number of insights:

\begin{itemize}
\item \emph{Performance bounds with ML detection:}
Recent results in
\cite{Wainwright:07-725,FletcherRG:08arXiv,WangWR:08arXiv}
provide simple upper and lower bounds on the
number of measurements required to detect the users reliably
assuming maximum likelihood (ML) detection.
One of the consequences of these bounds is that, unlike
the classic MAC channel, the sum rate achievable with
random access signaling can be strictly less than
the rate achievable with coordinated transmissions with the same
total power.

\item \emph{Potential gains over single-user detection:}
ML detection can be considered as a type of multiuser detection.
Current commercial designs, however, almost universally use simple
single-user detection (see, for example \cite{QiuHZ:01} for a typical WCDMA
design).
The single-user detection performance can be estimated by
bounds given in \cite{RauhutSV:08,FletcherRG:08arXiv}.
The bounds show that ML detection offers a potentially large gain
over single-user detection, particularly at high SNRs.
The gap at high SNRs can be explained by a certain \emph{self-noise} limit
experienced by single-user detection.

\item \emph{Lasso- and OMP-based multiuser detection and
near--far resistance:}
ML sparsity detection is a well-known
NP-hard problem \cite{Natarajan:95}.
However, there are practical, but suboptimal,
algorithms such as the orthogonal matching pursuit (OMP)
\cite{ChenBL:89,MallatZ:93,PatiRK:93,DavisMZ:94} and
``lasso" \cite{Tibshirani:96} methods
in sparse estimation
that can be used for multiuser detection methods
for the on--off random access channel.
In comparison to single-user detection, we show that
these methods can offer improved performance when the dynamic range
in received power levels is large.
This near--far resistance feature is similar to that
of standard MMSE multiuser detection in CDMA systems \cite{LupasV:90}.

\item \emph{Improved high SNR performance with power shaping:}
While both lasso and OMP offer
improvements over single-user detection, there is still a large gap
in the performance of these algorithms in comparison to ML detection
at high SNRs.
Specifically, at high SNRs, ML achieves a fundamentally different scaling
in the number of measurements required for reliable detection
than that required by lasso, OMP and single-user detection.

We show, however, that when accurate power control is available,
the ML scaling can be theoretically achieved with a simplified version
of OMP, which we call sequential OMP (SeqOMP).
The method is analogous to the classic
successive interference cancellation (SIC) method for the MAC channel.
Specifically, users are deliberately targeted at different received
power levels and then detected and cancelled out in descending order of power.

While SeqOMP shows significant gains over single-user detection,
for most practical problem sizes it does worse than standard OMP,
even without power shaping.
However, we show, at least by simulation, that
power shaping can improve the performance of OMP as well.
\end{itemize}

The connection between sparsity detection methods such as OMP
and the SIC technique for the MAC channel has also been observed
in the recent work of Jin and Rao~\cite{JinR:08}.
A related work by Wipf and Rao~\cite{WipfR:06}
also gave some empirical evidence
for the benefit of power shaping when used in conjunction with sparse
Bayesian learning algorithms.
Both the works \cite{JinR:08} and \cite{WipfR:06} are
discussed in more detail below.
The results in this paper
make the connections between sparsity detection and the random
access MAC channel more precise by giving concrete conditions
on the detectability of the sparsity pattern,
characterizing the optimal power shaping distribution, and
contrasting the classic MAC and on--off random access MAC capacities.

The remainder of the paper is organized as follows.
The setting is formalized in Section~\ref{sec:chanMod}.
In particular, we define all the key problem parameters.
Results that can be derived from existing necessary and
sufficient conditions for sparsity pattern recovery are then
presented in Section~\ref{sec:csAnalysis}.
We will see that there is a potentially-large performance gap
between single-user detection and the optimal ML detection.
Existing ``practical'' multiuser detection techniques
perform significantly better than single-user detection in that
they are near--far resistant.
However, their performance saturates at high SNRs, falling well short
of ML detection.
Section~\ref{sec:SOMP} presents a new detection algorithm,
sequential orthogonal matching pursuit (SeqOMP),
that has near--far resistance under certain assumptions on power control.
Furthermore, with optimal power shaping, it does not suffer
from saturation at high SNRs.
Numerical experiments are reported in Section~\ref{sec:sim}.
Connection to MAC capacity are discussed in Section~\ref{sec:capacity},
conclusions are given in Section~\ref{sec:concl},
and proofs are relegated to the Appendix.

\section{On-Off Random Access Channel Model}
\label{sec:chanMod}

\subsection{Problem Formulation}
Assume that there are $n$ transmitters sharing a wireless channel to a
single receiver.
Each user $j$ is assigned a unique, dedicated codeword represented as an 
$m$-dimensional vector $\abf_j \in \C^m$,
where $m$ is the total number of degrees of freedom in the channel.
By degrees of freedom we simply mean the dimension of the received
vector, which represents the number of samples in time or frequency
depending on the modulation.
In any channel use, only some fraction of the users, $\lambda \in (0,1)$,
transmit their codeword.
The fraction $\lambda$ will be called the \emph{activity ratio} and
any user that transmits will be called \emph{active}.

The signal at the receiver from each user $j$ is modeled as $x_j \abf_j$ where
$x_j$ is a complex scalar.
If the user is not active, $x_j = 0$.
If the user is active,
$x_j$ would represent the product of the transmitted symbol and channel gain.
The total signal at the receiver is given by
\beq \label{eq:yax}
    \ybf = \sum_{j=1}^{n} \abf_j x_j + \wbf = \Abf \xbf + \wbf,
\eeq
where $\wbf \in \C^m$ represents noise.  The matrix $\Abf \in \C^{m \times n}$
is formed by codewords $\abf_j$,
\[
    \Abf = \left[\abf_1 \ \cdots \abf_n\right],
\]
and will be called the \emph{codebook}.
The vector $\xbf = [x_1 \cdots x_n]^T$ will be called the
\emph{modulation vector},
and its components $\{x_j\}_{j=1}^n$ are referred to as the
\emph{received modulation symbols}.

Given a modulation vector $\xbf$, define the \emph{active user set} as
\beq\label{eq:Itrue}
    \Itrue = \left\{ ~j~: x_j \neq 0 ~\right\},
\eeq
which is the ``true" set of active users.
The size of the active user set is related to the activity ratio through
\beq \label{eq:lamDef}
    \lambda = \frac{1}{n}|\Itrue|.
\eeq
The goal of the receiver is to determine an estimate $\Ihat = \Ihat(\ybf)$
of $\Itrue$ based on the received noisy vector $\ybf$.

For the most part, we will be interested in estimators that exploit minimal
prior knowledge of the modulation vector $\xbf$ other than it being sparse.
In particular, we will limit our attention to estimators that do not
explicitly require \emph{a priori} knowledge of the
complex modulation symbols $x_j$.
This assumption is required since the channel gain is
typically unknown at the receiver in random access channels,
since users conducting random access communication would be
unlikely to be sending any other persistent pilot reference.

We consider large random codebooks where the entries of $\Abf$ are
i.i.d.\ ${\mathcal{CN}}(0,1/m)$.  We assume the noise vector is also
Gaussian: $\wbf \sim {\mathcal{CN}}(0,(1/m)I_m)$.
Given an estimator, $\Ihat = \Ihat(\ybf)$, the probability of error,
\beq \label{eq:Perr}
    \Perr = \Pr\left( \Ihat \neq \Itrue \right),
\eeq
is taken with respect to random codebook $\Abf$, the
noise vector $\wbf$, and the statistical distribution of the modulation vector $\xbf$.
We want to find estimators  $\Ihat$ that bring $\Perr$ close to zero.

We will see that two key factors influence the ability to detect the active user set.
The first is the total SNR defined as
\beq \label{eq:snrDef}
    \SNR = \frac{\Exp\|\Abf\xbf\|^2}{\Exp\|\wbf\|^2}.
\eeq
Since the components of the matrix $\Abf$ and noise vector $\wbf$ are i.i.d.\
${\mathcal{CN}}(0,1/m)$, it can be verified that, for deterministic $\xbf$,
\beq \label{eq:snrVal}
    \SNR = \|\xbf\|^2.
\eeq
In the case of random $\xbf$, this expression is the conditional SNR
given $\xbf$; we will have both deterministic and random formulations.

The second term is what we will call the
\emph{minimum-to-average ratio}
\beq \label{eq:MAR-def}
    \MAR = \frac{\min_{j \in \Itrue} |x_j|^2}
                {{\|\xbf\|^2}/{\lambda n}} .
\eeq
Since $\Itrue$ has $\lambda n$ elements,
$\|\xbf\|^2 / \lambda n$ is the average of $\{ |x_j|^2 \mid j \in \Itrue\}$.
Therefore, $\MAR \in (0,1]$ with the upper limit occurring when all the nonzero
entries of $\xbf$ have the same magnitude.
{\MAR} is a deterministic quantity when $\xbf$ is deterministic
and a random variable otherwise.

One final value that will be important is the \emph{minimum component SNR},
which, for a given $\xbf$, is given by
\beq \label{eq:snrMinDef}
    \SNRmin = \frac{1}{\Exp\|\wbf\|^2} \min_{j \in \Itrue} \Exp\|\abf_jx_j\|^2
    = \min_{j \in \Itrue} |x_j|^2,
\eeq
where $\abf_j$ is the $j$th column of $\Abf$.
The quantity $\SNRmin$ has a natural interpretation:
The numerator, $\min \Exp\|\abf_jx_j\|^2$ is the signal power due to
the smallest nonzero component in $\xbf$, while
the denominator, $\Exp \|\wbf\|^2$, is the total
noise power. The ratio $\SNRmin$ thus represents the contribution to
the SNR from the smallest nonzero component of the unknown vector $\xbf$.

The final equality in (\ref{eq:snrMinDef}) is a consequence of the fact that
$\Exp\|\abf_j\|^2  = \Exp\|\wbf_j\|^2 = 1$.
Observe that (\ref{eq:snrVal}) and (\ref{eq:MAR-def}) show
\beq \label{eq:snrMarProd}
    \SNRmin = \min_{j \in \Itrue} |x_j|^2
        = \frac{1}{\lambda n} \, \SNR \cdot \MAR .
\eeq

\subsection{MAR and Power Control} \label{sect:marPWC}

For wireless systems, the factor $\MAR$ in (\ref{eq:MAR-def})
has an important interpretation as a measure of the
dynamic range of received power levels.
With accurate power control, all users can be controlled to arrive at the same power.
In this case, $\MAR = 1$.
However, if power control is difficult due to fading
or lack of power control feedback, there can be a considerable dynamic range in the
received powers from different users.  In this case, some users could arrive
at powers much below the average making $\MAR$ closer to zero.

One of the results in this paper is a precise
quantification of the effect of MAR on the detectability of the active user set.
Specifically, we will show that low MAR can make reliable detection significantly more
difficult for certain algorithms.
The problem is analogous to the well-known \emph{near--far effect}
in CDMA systems \cite{LupasV:90},
where users with weak signals can be dominated by higher-power signals.

\subsection{Synchronization and Multi-Path}
It is important to recognize that an implicit assumption in the above model
is that the transmissions from different users are perfectly synchronized.
At a minimum, the timing offsets from the users are exactly
known at the receiver and there is no multipath.

Of course, in many wireless applications, exact synchronization is not possible
and the receiver must estimate the timing delay of the transmission
as part of the detection process.
In most practical receivers, timing offsets are estimated by discretizing
the delay search space, typically to a quarter or half-chip resolution.
The receiver then searches over a finite set of delay hypotheses depending on
the range of timing uncertainty.  In the presence of multipath, the receiver could
detect multiple delay hypotheses.

To model this search in the theoretical framework of this paper, we would
need to model each timing shift of the codeword as a different codeword.
The total number of codewords would then grow to the number of users times
the number of delay hypotheses per user.  While the algorithms we will present
can be applied in this manner to deal with the asynchronous case,
there are several theoretical issues with extending the analysis.
In particular, this extended codebook would lack the independence of
codewords that the simpler model has by construction.
We will thus just consider only the synchronous case for the remainder of
this paper.

\section{Performance with Current Sparsity Detection Methods}
\label{sec:csAnalysis}

The problem of detecting the active user set is precisely equivalent to
a sparsity pattern recovery problem.  To see this, note that the
modulation vector $\xbf$ is \emph{sparse}, with nonzero components only
in positions corresponding to the active users.  The problem at the
receiver is to detect these nonzero positions in $\xbf$ from noisy linear
observations $\ybf$ in (\ref{eq:yax}).

In this section, we develop asymptotic analyses for detection of
the active users based on previous results on sparsity pattern recovery.
We model $\xbf$ as deterministic, so the quantities $\lambda$, $\SNR$,
$\MAR$ and $\SNRmin$ are also deterministic.
Since our formulation allows simple translation of results from
\cite{Wainwright:07-725,FletcherRG:08arXiv,Wainwright:06,TroppG:07},
we state these translations without detailed justifications.
Several results are here adjusted by a factor of two because we
have complex, rather than real, measurements.

Our results are expressed as scaling laws on the number of measurements
for \emph{asymptotic reliable detection} of the active user set.
We define this as follows:
\begin{definition}
Suppose that we are given deterministic sequences $m = m(n)$
and $\xbf = \xbf(n) \in \C^n$ that vary with $n$.
For a given detection algorithm $\Ihat = \Ihat(\ybf)$,
we then define the probability of error $\Perr$ in (\ref{eq:Perr})
where the probability is taken over the randomness of the codebook $\Abf$
and the noise vector $\wbf$.
Given the number of measurements $m(n)$ and modulation vector $\xbf(n)$,
the probability of error will then simply be
a function of $n$.  We say that the detection algorithm achieves
\emph{asymptotic reliable detection} when $\Perr(n) \arr 0$.
\end{definition}

Table \ref{table:summary} summarizes the results from this section
and previews results from Section~\ref{sec:SOMP}.

\newlength{\tableVertA}
\setlength{\tableVertA}{1.5mm}
\newlength{\tableVertB}
\setlength{\tableVertB}{1mm}
\newlength{\tableVertC}
\setlength{\tableVertC}{-1.5mm}
\begin{table*}
 \begin{center}
  \begin{tabular}{|c||c|c|}
    \hline
           & finite $\captionSNR \cdot \captionMAR$ &
           $\captionSNR \cdot \captionMAR \rightarrow \infty$ \\ \hline \hline

& & \\[\tableVertC]
Necessary for ML
           & $m >   \frac{1}
                      {\tableMAR \cdot \tableSNR}  \lambda n \log(n(1-\lambda))$
           & $m > \lambda n$ \\[\tableVertA]
           & Fletcher \etal \cite[Thm.~1]{FletcherRG:08arXiv}
           & (elementary) \\[\tableVertB]
\hline

& & \\[\tableVertC]
Sufficient for ML
           & $m >   \frac{C}
                      {\tableMAR \cdot \tableSNR}  \lambda n \log(n(1-\lambda))$
           & $m > \lambda n$ \\[\tableVertA]
           & Wainwright \cite{Wainwright:07-725}
           & (elementary) \\[\tableVertB]
\hline

& & \\[\tableVertC]
Sufficient for sequential
           & $m > \frac{4}{\log(1+\tableSNR)}\lambda n \log(n(1-\lambda))$
           & $m > 5\lambda n$ \\[\tableVertA]
OMP with power shaping
           & {\bf From Theorem \ref{thm:minMeasSOMP}
             (Section~\ref{sec:powShaping})}
           & {\bf From Theorem \ref{thm:minMeasSOMP}
             (Section~\ref{sec:snrSaturation})} \\[\tableVertB]
\hline

& & \\[\tableVertC]
Necessary and
           & unknown (expression to
           & $m > \lambda n \log(n(1-\lambda))$ \\[\tableVertA]
sufficient for lasso
           & the right is necessary)
           & Wainwright~\cite{Wainwright:06} \\[\tableVertB]
\hline

& & \\[\tableVertC]
Sufficient for
           & unknown     & $m > 2\lambda n \log(n)$ \\[\tableVertA]
OMP
           &       & Tropp and Gilbert~\cite{TroppG:07} \\[\tableVertB]
\hline

& & \\[\tableVertC]
Sufficient for single
           & $m > \frac{4(1+\tableSNR)}
                       {\tableMAR \cdot \tableSNR}\lambda n \log(n(1-\lambda))$
           & $m > \frac{4}{\tableMAR}\lambda n \log(n(1-\lambda))$ \\[\tableVertA]
user detection (\ref{eq:IhatSUD})
           & Fletcher \etal \cite[Thm.~2]{FletcherRG:08arXiv} & \\[\tableVertB]
\hline

  \end{tabular}
 \end{center}
 \caption{Summary of results on measurement scalings for
   asymptotic reliable detection for various detection algorithms. \newline
   Only leading terms are shown.
   See body for definitions and additional technical limitations.}
 \label{table:summary}
\end{table*}

\subsection{Optimal Detection with No Noise}
\label{sec:noNoise}

To understand the limits of detection, it is useful to first
consider the minimum number of measurements when there is no noise.
Since the activity ratio is $\lambda$,
$\xbf$ will have $k = \lambda n$ nonzero components.
For a lower bound on the minimum number of measurements needed
for reliable detection, suppose that the receiver knows
the number of active users $k$ as side information.

With no noise, the received vector is $\ybf = \Abf \xbf$,
which will belong to one of $J = {n \choose k}$ subspaces
spanned by $k$ columns of $\Abf$.  If $m > k$,
then these subspaces will be distinct with probability 1\@.
Thus, an exhaustive search through
the subspaces will reveal which subspace $\ybf$ belongs to and thus
determine the active user set.
This shows that with no noise and no computational limits, the scaling
in measurements of
\beq \label{eq:minMeasOptNoNoise}
    m > \lambda n
\eeq
is sufficient for asymptotic reliable detection.

Conversely, if no prior information is known at the receiver
other than $\xbf$ being $k$-sparse, then the condition 
(\ref{eq:minMeasOptNoNoise}) is also necessary.
If $m \leq k = \lambda n$, then for almost all codebooks $\Abf$,
any $k$ columns of $\Abf$ span $\C^m$.  Consequently, any received
vector $\ybf = \Abf \xbf$ is consistent with any $k$ users transmitting.
Thus, the active user set cannot be determined without further
prior information on the modulation vector $\xbf$.

\subsection{ML Detection with Noise}
Now suppose there is noise.
Since $\xbf$ is an unknown deterministic quantity, the probability of
error in detecting the active user set is minimized by maximum likelihood
(ML) detection.
Since the noise $\wbf$ is Gaussian, the ML detector finds the $k$-dimensional
subspace spanned by $k$ columns of $\Abf$ containing the maximum energy of $\ybf$.

The ML estimator was first analyzed by Wainwright~\cite{Wainwright:07-725}.
The results in that work, along with the fact that $k = \lambda n$,
show that there exists a constant $C > 0$ such that if
\beqa
    m &\geq& C
        \max\left\{\frac{1}{\MAR \cdot \SNR}\lambda n \log(n(1-\lambda)),
        \lambda n\log(1/\lambda) \right\} \nonumber \\
      &=&C\max\left\{\frac{1}{\SNRmin}\log(n(1-\lambda)),
        \lambda n\log(1/\lambda) \right\}
      \label{eq:minMeasMLSuff}
\eeqa
then ML will asymptotically detect the correct active user set.
The equivalence of the two expressions in (\ref{eq:minMeasMLSuff}) is due to
(\ref{eq:snrMarProd}).
Also, \cite[Thm.~1]{FletcherRG:08arXiv}
(generalized in \cite[Thm.~1]{WangWR:08arXiv})
shows that, for any $\delta > 0$, the condition
\beqa
    m &\geq&
        \frac{1-\delta}{\MAR \cdot \SNR}\lambda n \log(n(1-\lambda))+ \lambda n,
            \nonumber \\
        &=&\frac{1-\delta}{\SNRmin} \log(n(1-\lambda))+ \lambda n,\label{eq:minMeasML}
\eeqa
is necessary.
Observe that when $\SNR \cdot \MAR \arr \infty$, the lower
bound (\ref{eq:minMeasML}) approaches $m \geq \lambda n$, matching the noise free
case (\ref{eq:minMeasOptNoNoise}) as expected.

These necessary and sufficient conditions for ML appear in
Table~\ref{table:summary} with smaller terms and the infinitesimal $\delta$
omitted for simplicity.

\subsection{Single User Detection} \label{sect:sud}
The most common and simple method to detect the active user set is a
single-user detection estimator of the form,
\beq \label{eq:IhatSUD}
    \Ihat_{SUD} = \left\{ ~j~: ~\rho(j) > \mu ~\right\},
\eeq
where $\mu > 0$ is a threshold parameter and $\rho(j)$ is the
correlation coefficient,
\beq \label{eq:rhoSUD}
    \rho(j) = \frac{|\abf_j'\ybf|^2}{\|\abf_j\|^2\|\ybf\|^2}.
\eeq
Single-user detection has been analyzed in the compressed sensing context
in \cite{DuarteSBWB:05,RauhutSV:08,FletcherRG:08arXiv}.
A small modification of \cite{FletcherRG:08arXiv}
shows the following result:  Suppose,
\beqa
    m(n) &>& \frac{(1 + \delta)L(\lambda,n)(1+\SNR)}{\SNR \cdot \MAR}\lambda n, \nonumber\\
    &=& \frac{(1 + \delta)L(\lambda,n)(1+\SNR)}{\SNRmin} \label{eq:minMeasSUDFull}
\eeqa
where $\delta > 0$ and
\beq \label{eq:Ldef}
    L(\lambda,n) = \left[\sqrt{\log(n(1-\lambda))} + \sqrt{\log(n\lambda)}\right]^2.
\eeq
Then there exists a sequence of detection thresholds $\mu = \mu(n)$
such that single-user detection achieves asymptotic reliable detection
of the active user set.
As before, the equivalence of the two expressions in (\ref{eq:minMeasSUDFull}) is due to
(\ref{eq:snrMarProd}).

Comparing the sufficient condition (\ref{eq:minMeasSUDFull})
for single-user detection with the \emph{necessary} condition (\ref{eq:minMeasML}),
we see two distinct problems in single-user detection:
\begin{itemize}
\item \emph{Constant offset:}  The scaling (\ref{eq:minMeasSUDFull})
for single-user detection shows a factor $L(\lambda,n)$ instead of
$\log((1-\lambda)n)$ in (\ref{eq:minMeasML}).
It is easily verified that, for $\lambda \in (0,1/2)$,
\beq \label{eq:lnlamBnd}
    \log((1-\lambda)n) < L(\lambda,n) < 4\log((1-\lambda)n),
\eeq
so this difference in factors alone could require that single-user detection
use up to four times more measurements than ML for asymptotic reliable detection.

Combining the inequality (\ref{eq:lnlamBnd}) with (\ref{eq:minMeasSUDFull}),
we see that the more stringent, but simpler, condition
\beq \label{eq:minMeasSUD}
    m(n) > \frac{(1 + \delta)4(1+\SNR)}{\SNR \cdot \MAR}
        \lambda n\log((1-\lambda)n)
\eeq
is also sufficient for asymptotic reliable detection with
single-user detection.  This simpler condition is shown in Table \ref{table:summary},
where we have omitted the infinitesimal $\delta$ quantity to simplify the
table entry.

\item \emph{Self noise limit:}  In addition to the $L(\lambda,n)/\log(n(1-\lambda)$
offset, single-user detection also requires a factor of
$1+\SNR$ more measurements than ML\@.
This $1+\SNR$ factor has a natural interpretation as \emph{self-noise}:
When detecting any one
component of the vector $\xbf$, single-user detection sees the energy from the
other $n-1$ components of the signal as interference.
We can think of this additional noise as self-noise, by which we mean
the interference caused from different components of the signal $\xbf$ interfering
with one another in the observed signal $\ybf$ through the measurement matrix $\Abf$.
This self-noise is distinct from the additive noise $\wbf$.
This self-noise increases
the effective noise by a factor of $1+\SNR$, which results in a proportional
increase in the minimum number of measurements.

This self-noise results in a large performance gap at high SNRs.
In particular, as $\SNR \arr \infty$, (\ref{eq:minMeasSUDFull}) reduces to
\beq \label{eq:minMeasSUDlim}
    m(n) > \frac{(1 + \delta)L(\lambda,n)}{\MAR}
        \lambda n\log((1-\lambda)n).
\eeq
In contrast, ML may be able to succeed obtain with a scaling $m = O(\lambda n)$ for high SNRs,
which is fundamentally better than the $m = \Omega(\lambda n\log((1-\lambda)n)$
required by single-user detection.
\end{itemize}

\subsection{Lasso and OMP Estimation}

While ML has clear advantages over single-user detection,
it is not computationally feasible.  However, one practical method
used in sparse signal estimation is the
\emph{lasso} estimator \cite{Tibshirani:96}, also called basis pursuit denoising
\cite{ChenDS:99}.
In the context of the random access channel,
the lasso estimator would first estimate the modulation vector $\xbf$
by solving the convex minimization
\beq \label{eq:xhatLasso}
    \xbfhat = \argmin_{\xbf} \left( \|\ybf - \Abf\xbf\|_2^2 + \mu\|\xbf\|_1 \right),
\eeq
where $\mu > 0$ is an algorithm parameter that ``encourages"
sparsity in the solution $\xbfhat$.
The nonzero components of $\xbfhat$ can then be used as an estimate of
the active user set.

The exact performance of lasso is not known at finite SNR\@.
However, Wainwright \cite{Wainwright:06} has exactly characterized the
conditions for lasso to work in the high SNR regime.  Specifically,
if $m$, $n$ and $\lambda n \arr \infty$, with $\SNR \cdot \MAR \arr \infty$,
the scaling
\beq \label{eq:minMeasLasso}
    m > \lambda n \log(n(1-\lambda)) + \lambda n + 1,
\eeq
is both necessary and sufficient for asymptotic sparsity recovery.

Another common approach to sparsity pattern detection is the greedy
OMP algorithm~\cite{ChenBL:89,PatiRK:93,DavisMZ:94}.
This has been analyzed by Tropp and Gilbert~\cite{TroppG:07}
in a setting with no noise.
They show that, when $\Abf$ has Gaussian entries,
a \emph{sufficient} condition for asymptotic reliable recovery is
\beq \label{eq:minMeasOMP}
    m > 2\lambda n\log(n) + C\lambda n,
\eeq
where $C > 0$ is a constant.
Numerical experiments reported in~\cite{TroppG:07} suggest that
the constant factor $2$ may be removed,
although this has not be proven.
In any case, OMP with no noise has a similar scaling in the sufficient
number of measurements as lasso.

The conditions (\ref{eq:minMeasLasso}) and (\ref{eq:minMeasOMP})
are both shown in Table~\ref{table:summary}.
As usual, the table entries are simplified by including only
the leading terms.

The lasso and OMP scaling laws, (\ref{eq:minMeasLasso}) and (\ref{eq:minMeasOMP}),
can be compared with the high SNR limit for the single-user detection
scaling law in (\ref{eq:minMeasSUDlim}).  This comparison shows the following:

\begin{itemize}
\item \emph{Removal of the constant offset:}
The $L(\lambda,n)$ term in the single-user detection expression (\ref{eq:minMeasSUDlim})
is replaced by a $\log(n(1-\lambda))$ term in the lasso scaling law
(\ref{eq:minMeasLasso}) and $2\log(n)$ for the OMP scaling law (\ref{eq:minMeasOMP}).
Similar to the discussion above, this implies that lasso could require up
to 4 times fewer measurements than single-user detection.  OMP could
require 2 times fewer.

\item \emph{Near--far resistance:}
In addition, both the lasso and OMP methods do not have a dependence
on MAR; thus, in the high SNR regime, they have a near--far resistance that
single-user detection does not.  This gain can be large when there are
users whose received powers are much below the average (low MAR).

The near--far resistance of lasso and OMP is analogous to that of MMSE
multiuser detection in CDMA systems \cite{LupasV:90}.
In that case, when the number of degrees of freedom $m$ exceeds the number
of users $n$, a decorrelating detector can null out strong users while recovering
weak ones.  An interesting property that we see in the random access case
is that near--far resistance may be possible when $m < n$, provided that
$m$ is sufficiently greater than the number of $\emph{active users}$, $\lambda n$.

\item \emph{Limits at high SNR:}
We also see from (\ref{eq:minMeasLasso}) and
(\ref{eq:minMeasOMP}) that both lasso and OMP are unable to achieve
the scaling $m = O(\lambda n)$ that may be achievable with ML at high SNR\@.
Instead, both lasso and OMP have the scaling, $m = O(\lambda n \log((1-\lambda)n))$,
similar to the minimum scaling possible with single-user detection,
which suffers from a self-noise limit.
\end{itemize}

\subsection{Other Sparsity Detection Algorithms}
Recent interest in compressed sensing has led to a plethora
of algorithms beyond OMP and lasso.
Empirical evidence suggests that the most promising algorithms
for sparse pattern detection are the sparse Bayesian learning methods
developed in the machine learning community in \cite{Tipping:01},
and introduced into signal processing applications in \cite{WipfR:04},
with related work in \cite{SchniterPZ:0x}.
Unfortunately, a comprehensive summary of these
algorithms is far beyond the scope of
this paper.

Instead,  we will limit our discussion to the lasso and OMP methods since these
are the algorithms with the most concrete analytic results
on asymptotic reliable detection.
Moreover, our interest is not in finding the optimal algorithm,
but merely to point out general qualitative effects such
as near--far and self-noise limits which should be considered
in evaluating any algorithm.

\section{Sequential Orthogonal Matching Pursuit}
\label{sec:SOMP}

The analyses in the previous section suggest that ML detection may offer significant
gains over the provable performance of current ``practical'' algorithms such as
single-user detection, lasso and OMP, when the SNR is high.
Specifically, as the SNR increases, the performance of these practical
methods saturates at a scaling in the number of measurements that
can be significantly higher than that for ML\@.

In this section, we show that if accurate power control is available,
an OMP-like algorithm, which we call
\emph{sequential orthogonal matching pursuit} or SeqOMP, can break this
barrier.  Specifically, the performance of SeqOMP does
not saturate at high SNR.

\subsection{Algorithm} \label{sect:somp}
\begin{algorithm}[SeqOMP]
\label{algo:somp}  Given a received vector $\ybf$
and threshold level $\mu > 0$,
the algorithm produces an estimate $\IhatSOMP$ of the active
user set with the following steps:
\begin{enumerate}
\item Initialize the counter $j=1$ and
set the initial active user set estimate to empty:  $\Ihat(0) = \{\emptyset\}$.
\item Compute $\Pbf(j)\abf_j$
where $\Pbf(j)$ is the projection operator
onto the orthogonal complement
of the span of $\{ \abf_\ell, \ell \in \Ihat(j-1)\}$.
\item Compute the correlation,
\beq \label{eq:rhoSOMP}
    \rho(j) = \frac{|\abf_j'\Pbf(j)\ybf|^2 }{\|\Pbf(j)\abf_j\|^2\|\Pbf(j)\ybf\|^2}.
\eeq
\item If $\rho(j) > \mu$, add the index $j$ to $\Ihat(j-1)$.  That is,
$\Ihat(j) = \Ihat(j-1) \cup \{j\}$.
Otherwise, set $\Ihat(j) = \Ihat(j-1)$.
\item Increment $j = j +1$.  If $j \leq n$ return to step 2.
\item The final estimate of the active user set is $\IhatSOMP = \Ihat(n)$.
\end{enumerate}
\end{algorithm}

The SeqOMP algorithm can be thought of as an iterative version of single-user
detection with the difference that, after an active user
is detected, subsequent correlations are performed only in the orthogonal complement
to the detected codeword.  The method
is identical to the standard OMP algorithm of
\cite{ChenBL:89,PatiRK:93,DavisMZ:94},
except that SeqOMP passes through the data only once.
For this reason, SeqOMP is actually computationally simpler than standard OMP\@.

As simulations will illustrate later, SeqOMP generally has much worse
performance than standard OMP\@.  It is not intended as a competitive practical
alternative.  Our interest in the algorithm lies in the fact
that we can prove positive results for SeqOMP\@.
Specifically, we will be able to show that this relatively poor algorithm,
when used in conjunction with \emph{power shaping},
can achieve a fundamentally better scaling at high SNRs
than what has been proven is achievable with methods such as OMP\@.
We will also provide some simulation evidence that OMP can also benefit somewhat
from power shaping, although we will not be able to prove this here.

\subsection{Sequential OMP Performance}
The analysis in Section \ref{sec:csAnalysis} was based on deterministic
vectors $\xbf$.
To characterize the SeqOMP performance, it is simpler to use
a partially-random model
where the active user set is random while the received modulation
signal power $|x_j|^2$, conditioned on user $j$ being active,
remains deterministic.
We reuse the notation $\lambda$ because its meaning remains almost the same.

We assume that each user is active with some probability $\lambda \in (0,1)$,
which we now call the \emph{activity probability}.  The activities of different
users are assumed to be independent.
Thus, unlike in Section~\ref{sec:csAnalysis}, $\lambda n$ represents
the \emph{average} number of users that are active, as opposed to the actual number.

Let $p_j$ denote the received modulation symbol power
\beq \label{eq:pjdef}
    p_j = |x_j|^2,
\eeq
conditional that user $j$ is active.
We will call the set $\{p_j\}_{j=1}^n$ the \emph{power profile},
which we will treat as a deterministic quantity.
Since each user transmits with a probability $\lambda$, the total average
SNR is given by,
\beq \label{eq:pellTotCon}
    \SNR = \lambda \sum_{\ell=1}^n p_j.
\eeq
This factor is also deterministic.

Given a power profile, we will see that a key parameter in estimating
the performance of the SeqOMP algorithm is what we will call
the \emph{minimum signal-to-interference and noise ratio (SINR)} defined as
\beq \label{eq:sinrDef}
    \gamma = \min_{\ell=1,\ldots,n}  p_\ell / \sigmahat^2(\ell),
\eeq
where $\sigmahat^2(\ell)$ is given by
\beq \label{eq:sigHatDef}
    \sigmahat^2(\ell) = 1 + \lambda \sum_{j=\ell+1}^n p_j.
\eeq
The parameters $\gamma$ and $\sigmahat^2(\ell)$ have simple interpretations:
Suppose that the SeqOMP algorithm
has correctly decoded all the users for $j < \ell$.
Then, in detecting the $\ell$th user, the receiver sees the noise $\wbf$
with power $\Exp \|\wbf\|^2 = 1$ and, for each user $j > \ell$, an
interference power $p_j$ with probability $\lambda$.
Hence, $\sigmahat^2(\ell)$ is the total average interference power seen when
detecting $\ell$th user, assuming perfect cancellation.
Since user $\ell$ arrives at a power $p_\ell$, the ratio $p_\ell / \sigmahat^2(\ell)$
in (\ref{eq:sinrDef}) represents the average SINR seen by user $\ell$.
The value $\gamma$ is the minimum SINR over all $n$ users.

\medskip

\begin{theorem} \label{thm:minMeasSOMP}
Let $\lambda = \lambda(n)$, $m = m(n)$ and
the power profile $\left\{p_j\right\}_{j=1}^n = \left\{p_j(n)\right\}_{j=1}^n$,
be deterministic quantities that all vary with $n$
satisfying the limits $m-\lambda n$, $\lambda n$ and
$(1- \lambda) n \arr \infty$, and $\gamma \arr 0$.
Also, assume the sequence of power profiles satisfies the limit
\beq \label{eq:pjsqbnd}
    \limn \max_{i=1,\ldots,n-1} \log(n)\sigmahat^{-4}(i)\sum_{j>i}^n p_j^2 = 0.
\eeq
Finally, assume that for all $n$,
\beq \label{eq:minMeasGam}
    m \geq \frac{(1 + \delta)L(n,\lambda)}{\gamma} + \lambda n,
\eeq
for some $\delta > 0$ and $L(n,\lambda)$ defined in (\ref{eq:Ldef}).
Then, there exists a sequence of thresholds, $\mu = \mu(n)$,
such that SeqOMP will achieve asymptotic reliable detection of the active user set
in that
\[
    \Perr = \Pr\left( \IhatSOMP \neq \Itrue \right) \arr 0,
\]
where the probability is taken over the randomness in the activities of the users,
the codebook $\Abf$, and the noise $\wbf$.
The sequence of threshold levels can be selected independent of the
sequence of power profiles.
\end{theorem}
\begin{proof} See Appendix~\ref{sec:proofOutline}.
\end{proof}

The theorem provides a simple sufficient condition on the number of
measurements as a function of the SINR $\gamma$, activity probability $\lambda$
and number of users $n$.  The condition (\ref{eq:pjsqbnd})
is somewhat technical, but is satisfied in the cases that interest us.
The remainder of this section will discuss some of the implications of this theorem.

\subsection{Near--Far Resistance with Known Power Ordering}
\label{sec:nearFarSOMP}

First, suppose that the power ordering $p_j$ is known at the receiver
so the receiver can detect the users in order of decreasing power.
If, in addition, the SNRs of all the users go to infinity so that $p_j \arr \infty$
for all $j$, then it can be verified that $\gamma > 1/(\lambda n)$.
In this case, the sufficiency of the scaling (\ref{eq:minMeasGam})
shows that
\[
    m \geq (1 + \delta)\lambda n L(n,\lambda) + \lambda n
\]
is sufficient for asymptotic reliable detection.
This is identical to the lasso performance except for the factor
$L(\lambda,n) / \log((1-\lambda)n)$, which lies in $(0,4)$ for
$\lambda \in (0,1/2)$.
In particular, the minimum number of measurements does not depend on $\MAR$;
therefore, similar to lasso and OMP, SeqOMP can theoretically detect users
even when they are much below the average power.

With SeqOMP, simply knowing the order of powers is sufficient
to achieve near--far resistance when the SNR is sufficiently high.
Unlike for single-user detection,
unequal received powers do not hurt the performance of SeqOMP,
as long as the order of the powers are known at the receiver.
The feasibility of knowing the power ordering is addressed in
Section~\ref{sec:powerControl} below.
We will now look at the effect of the power profile on the performance.

\subsection{Performance with Constant Power}

Consider the case when all the powers $p_j$ are equal.
To satisfy the constraint (\ref{eq:pellTotCon}), the constant power level must be
$p_j = \SNR/(\lambda n)$.
From (\ref{eq:sinrDef}), the minimum SINR is $\gamma = \gammaConst$, where
\beq \label{eq:gamConst}
    \gammaConst = \frac{\SNR}{\lambda (n+(n-1)\SNR)} \approx \frac{\SNR}{\lambda n(1+\SNR)},
\eeq
and the approximation holds for large $n$.

It can be verified that the constant power profile satisfies
the technical condition (\ref{eq:pjsqbnd})
provided $\lambda$ is bounded away from zero and the SNR does not grow ``too fast".
Specifically, the SNR must satisfy $\SNR = o(n/\log(n))$.
In this case, we can substitute $\gamma = \gammaConst$ in (\ref{eq:minMeasGam})
to obtain the condition
\[
    m > \frac{(1+\delta)(1 + \SNR)L(\lambda,n)}{\SNR}\lambda n + \lambda n
\]
for asymptotic reliable detection.
The condition is precisely the condition for single-user detection in
(\ref{eq:minMeasSUD}) with MAR = 1 and an additional $\lambda n$ term.

Thus, for a constant power profile, Theorem \ref{thm:minMeasSOMP} does not
show any benefit in using SeqOMP\@.

\subsection{Optimal Power Shaping}
\label{sec:powShaping}

The constant power profile, however, is not optimal.
Suppose that accurate power control is feasible so that the
receive power levels $p_j$ can be set by the receiver.  In this case, we can
maximize the SINR $\gamma$ in (\ref{eq:sinrDef}) for a given
total SNR constraint (\ref{eq:pellTotCon}).
It is easily verified that any power profile $p_j$ maximizing the SINR $\gamma$
in (\ref{eq:sinrDef}) will satisfy
\beq \label{eq:pellSnrCon}
    p_\ell = \gamma\left( 1 + \lambda \sum_{j=\ell+1}^n p_j \right)
\eeq
for all $\ell = 1,\ldots,n$.
The solution to (\ref{eq:pellSnrCon}) and (\ref{eq:pellTotCon})
is given by
\beq \label{eq:pexp}
    p_\ell = \gamma(1+\lambda \gamma)^{n-\ell},
\eeq
where $\gamma = \gammaOpt$ is the SINR,
\beq \label{eq:gamExp}
    \gammaOpt = \frac{1}{\lambda}\left[(1 + \SNR)^{1/n} - 1\right]
    \approx \frac{1}{\lambda n}\log(1+\SNR).
\eeq
Here, the approximation holds for large $n$.
Again, some algebra shows that, when $\lambda$ is bounded away from zero,
the power profile $p_j$ in (\ref{eq:pexp}) will
satisfy the technical condition (\ref{eq:pjsqbnd}) when $\log(1+\SNR) = o(n/\log(n))$.

The power profile (\ref{eq:pexp}) is exponentially decreasing in the index order $\ell$.
Thus, users early in the detection sequence are allocated exponentially higher
power than users later in the sequence.  This allocation insures that
early users have sufficient power to overcome the interference
from all the users later in the detection sequence that are not yet cancelled.
This power shaping is analogous to the optimal power allocations in the classic
MAC channel when using a SIC receiver \cite{CoverT:91}.

The ratio of the optimal SINR $\gammaOpt$ in (\ref{eq:gamExp})
to the SINR with a constant
power profile, $\gammaConst$ in (\ref{eq:gamConst}) is given by
\[
    \frac{\gammaOpt}{\gammaConst} = \frac{(1+\SNR)\log(1+\SNR)}{\SNR}.
\]
This ratio represents the potential increase in SINR with exponential power shaping
relative to the SINR with equal power for all users.  The ratio increases with SNR and
can be large when the SNR is high.
For example, when $\SNR = 10\,\dB$, $\gammaOpt/\gammaConst \approx 2.6$.
When $\SNR = 20\,\dB$, the gain is even higher at
$\gammaOpt/\gammaConst \approx 4.7$.

Based on Theorem \ref{thm:minMeasSOMP}, this gain in SINR will result
in a proportional decrease in the minimum number of measurements.
Specifically, if we substitute the SINR $\gammaOpt$ in (\ref{eq:gamExp}) into
(\ref{eq:minMeasGam}), we see that that the condition
\beq \label{eq:minMeasSOMP}
    m \geq \frac{(1 + \delta)L(n,\lambda)}{\log(1+\SNR)}\lambda n + \lambda n
\eeq
is sufficient for SeqOMP to achieve asymptotic reliable detection of the active users,
when the users use exponential power shaping (\ref{eq:pexp}).

As before, if $\lambda < 1/2$, we can bound $L(n,\lambda) < 4\log(n(1-\lambda)$ and
the sufficient condition (\ref{eq:minMeasSOMP}) can be simplified to
\beq \label{eq:minMeasSMPSimp}
    m \geq \frac{4(1 + \delta)\log(n(1-\lambda))}{\log(1+\SNR)}\lambda n + \lambda n,
\eeq
the leading term of which appears in Table \ref{table:summary}
with the $\delta$ omitted.

\subsection{SNR Saturation}
\label{sec:snrSaturation}
As discussed earlier, a major problem with both single-user detection
and lasso multiuser detection was that their performance ``saturates" with high SNR\@.
That is, even as the SNR scales to infinity, the minimum number of measurements
scales as $m = O(\lambda n\log((1-\lambda)n)$.
In contrast, optimal ML detection can achieve a scaling $m = O(\lambda n)$,
when the SNR is sufficiently high.

An important consequence of (\ref{eq:minMeasSOMP}) is that SeqOMP
with exponential power shaping can overcome this bound.
Specifically, if we take the scaling of $\SNR = \Theta(\lambda n)$ in
(\ref{eq:minMeasSMPSimp}) and assume that $\lambda$ is bounded away from zero
we see that asymptotically, SeqOMP requires only
\beq \label{eq:minMeasSMPHigh}
    m \geq 5\lambda n
\eeq
measurements.
In this way, unlike single-user and lasso detection,
SeqOMP is able to obtain the scaling $m = O(\lambda n)$ when
the $\SNR \arr \infty$.

\subsection{Power Shaping with Sparse Bayesian Learning}
The fact that power shaping can provide benefits when
combined with certain iterative detection algorithms confirms the
observations in the work of Wipf and Rao~\cite{WipfR:06}.
That work considers signal detection with
a certain sparse Bayesian learning (SBL) algorithm.
They show the following result: Suppose $\xbf$ has $k$ non-zero
components and $p_i$, $i=1,2,\ldots,k$,
is the power of the $i$th largest component.
Then, for a given measurement matrix $\Abf$,
there exist constants $\nu_i > 1$ such that if
\beq \label{eq:sblPow}
    p_i \geq \nu_i p_{i-1},
\eeq
the SBL algorithm will correctly detect the sparsity pattern of $\xbf$.

The condition (\ref{eq:sblPow}) shows that a certain
growth in the powers can guarantee correct detection.
The parameters $\nu_i$ however depend in some complex manner on
the matrix $\Abf$,
so the appropriate growth is difficult to compute.
They also provide strong empirical evidence that shaping
the power with certain profiles can greatly reduce the number of
measurements needed.

The results in this paper add to Wipf and Rao's observations
showing that growth in the powers can also assist sequential OMP\@.
Moreover, for the SeqOMP case, we can explicitly derive the
optimal power profile for certain large random matrices.

This is not to say that SeqOMP is better than SBL\@.  In fact, 
empirical results in \cite{WipfR:04} suggest that SBL will outperform
OMP, which will in turn do better than SeqOMP\@.  As we have stressed before,
the point here of analyzing SeqOMP is that we can easily derive
concrete analytic results.  These results may provide guidance for
more sophisticated algorithms.

\subsection{Robust Power Shaping}
\label{sec:robustPowerShaping}
The above analysis shows certain benefits of SeqOMP used in conjunction with power shaping.
However, these gains are theoretically only possible at infinite block lengths.
Unfortunately, when the block length is finite, power shaping can actually
reduce the performance.

The problem is that when an active user is not detected in SeqOMP,
the user's energy is not cancelled out and remains as
interference for all subsequent users in the detection sequence.
With power shaping, users early in the detection sequence have much higher
power than users later in the sequence, so missing an early user can make the detection
of subsequent users difficult.  At infinite block lengths,
the probability of missing an active user can be driven to zero.  But,
at finite block lengths, the probability of missing an active user early
in the sequence will always be nonzero, and therefore
a potential problem with power shaping.

The work \cite{AgrawalACM:05} observed a similar problem when SIC is used in the
CDMA uplink.
To mitigate the problem, \cite{AgrawalACM:05} proposed to adjust
the power allocations to make them more robust to decoding errors early in the
decoding sequence.
The same technique,  which we will call \emph{robust power shaping},
can be applied to the SeqOMP as follows.

In the condition (\ref{eq:pellSnrCon}), it is assumed that all the energy
of users with index $j < \ell$ have been correctly detected and subtracted.
But, following \cite{AgrawalACM:05}, suppose that on average some fraction
$\theta \in [0,1]$ of the energy
of users early in the detection sequence is not cancelled out due to
missed detections.  We will call $\theta$ the \emph{leakage fraction}.
With nonzero leakage, the condition (\ref{eq:pellSnrCon})
would be replaced by
\beq \label{eq:pellSnrConTheta}
    p_\ell = \gamma \left( 1 + + \theta \lambda \sum_{j=1}^{\ell-1} p_j
        +\lambda \sum_{j=\ell+1}^n p_j \right).
\eeq
For given SNR, $\theta$ and $\lambda$, the linear equations
(\ref{eq:pellTotCon}) and (\ref{eq:pellSnrConTheta}) can be  solved
to obtain the optimal power profile, given by
\beq \label{eq:powTheta}
    p_j = \frac{(1-\theta)\gamma}{1+\lambda\theta\gamma}
    \left(\frac{1+\lambda\gamma}{1+\lambda\theta\gamma}\right)^{n-j},
\eeq
where $\gamma = \gamma(\theta)$, the optimal SINR
\beqa
    \gamma(\theta) &=& \frac{1}{\lambda(1-\theta)}\left[
        \left(\frac{1+\SNR}{1+\theta \, \SNR}\right)^{1/n} - 1\right] \nonumber \\
        &\approx& \frac{1}{\lambda n(1-\theta)}\log
        \left(\frac{1+\SNR}{1+\theta \, \SNR}\right). \label{eq:gamTheta}
\eeqa
The approximation here is valid for large $n$.

\begin{figure}
 \begin{center}
  \epsfig{figure=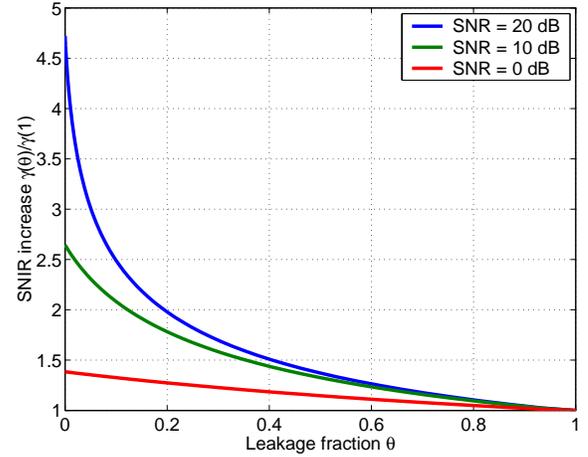,width=3in}
 \end{center}
 \caption{Increase in SINR, $\gamma(\theta)/\gammaConst = \gamma(\theta)/\gamma(1)$,
 as a function of the leakage fraction $\theta$.}
 \label{fig:sinrIncrease}
\end{figure}

Fig.\ \ref{fig:sinrIncrease} plots the SINR, $\gamma(\theta)$,
as a function of the leakage fraction $\theta$.  The SINR is plotted
relative to $\gammaConst$ in (\ref{eq:gamConst}), which is the SINR
that one obtains with a constant power profile.
The increase in SINR is maximized when the leakage fraction, $\theta = 0$.
When $\theta = 0$, $\gamma(\theta) = \gamma_{exp}$, the SINR (\ref{eq:gamExp})
for the exponential power shaping.  This is the optimal SINR, but assumes that
there are no missed detections.

As the leakage fraction $\theta$ is increased, the SINR, $\gamma(\theta)$,
decreases, which is price for the robustness to missed detections.
In the limit as $\theta \arr 1$, the optimal power profile, $p_j$ in
(\ref{eq:powTheta}) approaches a constant
and the corresponding SINR, $\gamma(\theta)$, converges to  $\gammaConst$.
However, even at a reasonable leakage fraction, say $\theta = 0.1$,
the SINR $\gamma(\theta)$ can still be significantly larger than $\gammaConst$.

\begin{figure}
 \begin{center}
  \epsfig{figure=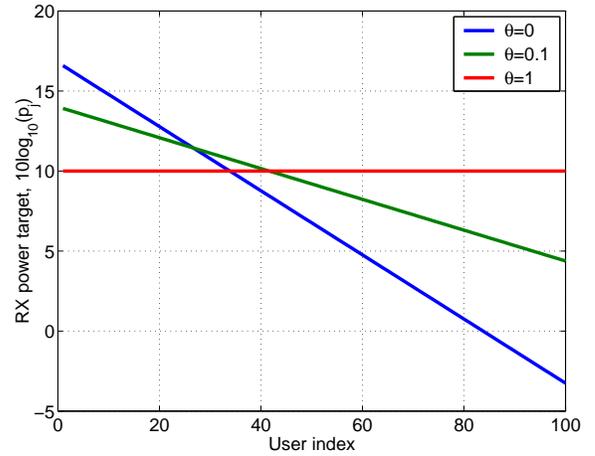,width=3in}
 \end{center}
 \caption{Robust power shaping:  Power profiles for various
 leakage values $\theta$.  For all the curves,
 the number of users is $n=100$, $\captionSNR = 20\,\dB$,
 and activity probability is $\lambda = 0.1$.}
 \label{fig:robPowProfile}
\end{figure}

It is illustrative to actually look at the optimal power profiles as a
function of $\theta$.
Fig.\ \ref{fig:robPowProfile} plots the optimal power profile, $p_j$ in
(\ref{eq:powTheta}), for leakage values
of $\theta =$ 0, 0.1 and 1.  In the plot, $n=100$,  $\SNR = 20\,\dB$, and
$\lambda = 0.1$.
It can be seen that when $\theta = 0$, there is a large range of almost
20 dB in the target receive powers from the first to last user.
While this power profile it optimal when there are no missed detections,
the power allocations can be very damaging if an active user is missed.
In an extreme case, for example, if the first user is active but not detected
and not cancelled it will cause
an interference level 20 dB above the signal level of the last user.
As the leakage fraction $\theta$ is increased, the range of powers is decreased,
which improves the robustness to missed detection at the expense of
reduced SINR\@.

\subsection{Practical Power Control Considerations}
\label{sec:powerControl}

In the original description of the problem in Section \ref{sec:chanMod},
we said that we would restrict our attention to estimators that
do not require \emph{a priori} knowledge of the modulation vector $\xbf$.
However, although SeqOMP does not require knowledge at the receiver
of the channel phases, the above analysis shows that 
knowledge of the \emph{order} of the conditional received powers
is necessary to achieve near--far resistance.
Additionally, eliminating the self-noise limit requires that powers are
explicitly targeted to a certain profile.

The use of power control for on--off random access communication
requires some justification.  On--off random access signaling is most
likely to be used when the users do not already have some ongoing
communication.  For example, in cellular systems, it is used
for initial access or requests to transmit.
If the users were already transmitting, the one bit could be embedded in the
other communication and on--off random access signaling would not be needed.
Consequently, fast feedback power control 
would likely not be available for such on--off random access transmissions
since the users are not likely to have a continuous transmission
to measure the received power.

Thus, in practice, power control is likely achievable only by
open-loop methods.  Open-loop power control is used for example
in cellular systems where each mobile estimates the path loss in the downlink
and adjusts its access power appropriately in the uplink.
Open-loop power control is most accurate when the uplink
and downlink are time-division duplexed (TDD) in the same band.  

\section{Numerical Simulation}
\label{sec:sim}

\subsection{Threshold Settings}
The performance of the single-user detection and SeqOMP algorithms
depend on the setting of the threshold level $\mu$.
In the theoretical analysis of Theorem \ref{thm:minMeasSOMP},
an ideal threshold is calculated assuming infinite
block lengths that guarantees perfect detection of the active user set.
However, in simulations with finite block lengths, it is more reasonable to set the
threshold based on a desired \emph{false alarm probability}.
A false alarm is the event when the algorithm falsely detects that a user is active
when it is not.  For the single-user detection algorithm in
Section \ref{sect:sud} or the SeqOMP algorithm in Section \ref{sect:somp},
the false alarm probability is
\beqan
    \PFA &=& \Pr\left( j \in \Ihat \mid j \not \in \Itrue \right) \\
    &=& \Pr\left( \rho(j) > \mu \mid j \not \in \Itrue \right),
\eeqan
which is the probability that the correlation $\rho(j)$ exceeds the threshold $\mu$
when the user $j$ is not active.

It is shown in the proof of Theorem~\ref{thm:minMeasSOMP} that,
when $j \not \in \Itrue$,
$\rho(j)$ follows a Beta $B(2,2(m-1))$ distribution.
When $m$ is large, this beta distribution is approximately Rayleigh and
the false alarm probability is given by
\[
    \PFA \approx \exp(-\mu m).
\]
Thus, the threshold level $\mu$ can be set to
\[
    \mu = -\log(\PFA) / m
\]
for a given desired false alarm probability.

In the simulations below, we will run the algorithms with a fixed false
alarm probability (typically $\PFA = 10^{-3}$), and measure
the \emph{missed detection rate} given by
\[
    \PMD = \Pr\left( j \not \in \Ihat \mid j \in \Itrue \right).
\]
The missed detection rate will be averaged over all $j \in \Itrue$.

\subsection{Evaluation of Bounds}
\begin{figure*}
 \begin{center}
  \epsfig{figure=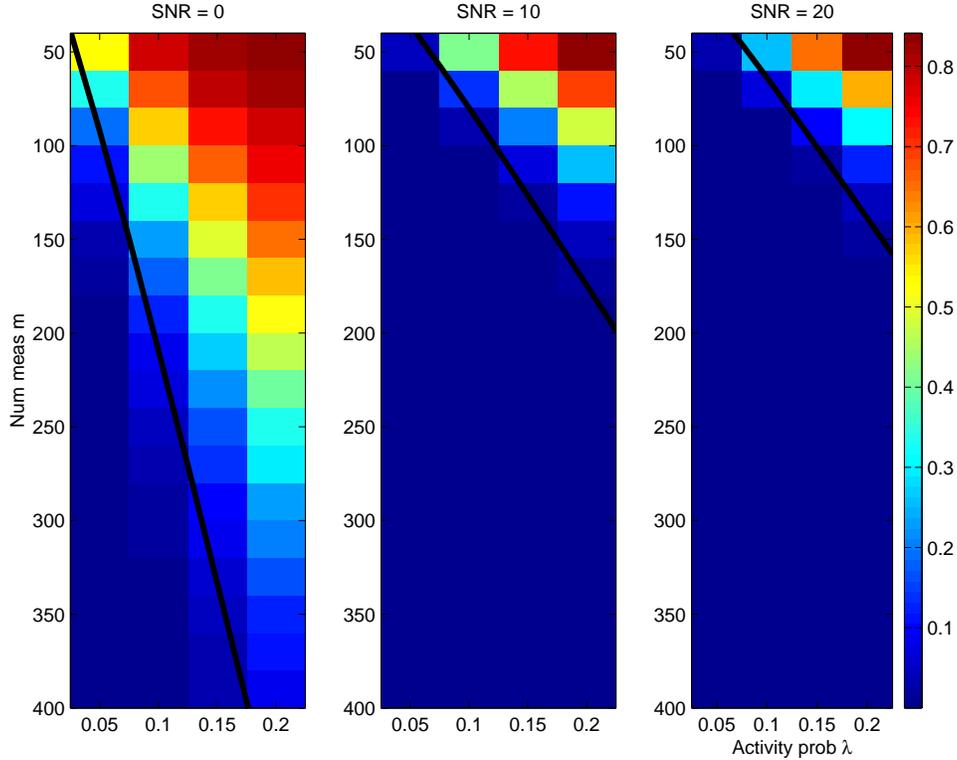,width=5in}
 \end{center}
 \caption{SeqOMP with power shaping:
 Each colored bar represents
 the SeqOMP algorithm's missed detection probability
 as a function of the number of measurements $m$,
 with different bars showing different activity probabilities $\lambda$ and
 SNR levels.  The missed detection probabilities were estimated with 1000 Monte Carlo
 trials.  The number of users is set to $n=100$, the false alarm
 probability is  $\PFA = 10^{-3}$.
 The power shaping is performed with a leakage fraction of $\theta=0.1$.
 The dark black line shows the theoretical number of measurements
 $m$ required in Theorem \ref{thm:minMeasSOMP} with $\gamma = \gamma(\theta)$
 in (\ref{eq:gamTheta}). }
 \label{fig:sompBnd}
\end{figure*}

We first compare the actual performance of the SeqOMP algorithm
with the bound in Theorem \ref{thm:minMeasSOMP}.
Fig.\ \ref{fig:sompBnd} plots the simulated missed detection probability for
using SeqOMP at various SNR levels, activity probabilities $\lambda$,
and numbers of measurements $m$.  In all simulations, the number of users was
fixed to $n=100$ and the users arrived at equal power ($\MAR = 1$).
The false alarm probability was set to $\PFA = 10^{-3}$.
The robust power profile of Section
\ref{sec:robustPowerShaping} is used with a leakage fraction $\theta = 0.1$.

The dark line in Fig.~\ref{fig:sompBnd} represents the number of measurements
$m$ for which Theorem~\ref{thm:minMeasSOMP} would theoretically guarantee reliable
detection of the active user set at infinite block lengths.
To apply the theorem, we used the SINR $\gamma = \gamma(\theta)$ in (\ref{eq:gamTheta}).
At the block lengths considered in this simulation,
the missed detection probability at the theoretical sufficient condition is small,
typically between 2 and 10\%.
Thus, even at moderate block lengths, the theoretical bound
in Theorem \ref{thm:minMeasSOMP} can provide a good estimate for the number of
measurements for reliable detection.

\subsection{SeqOMP vs.\ Single User Detection}

\begin{figure}
 \begin{center}
  \epsfig{figure=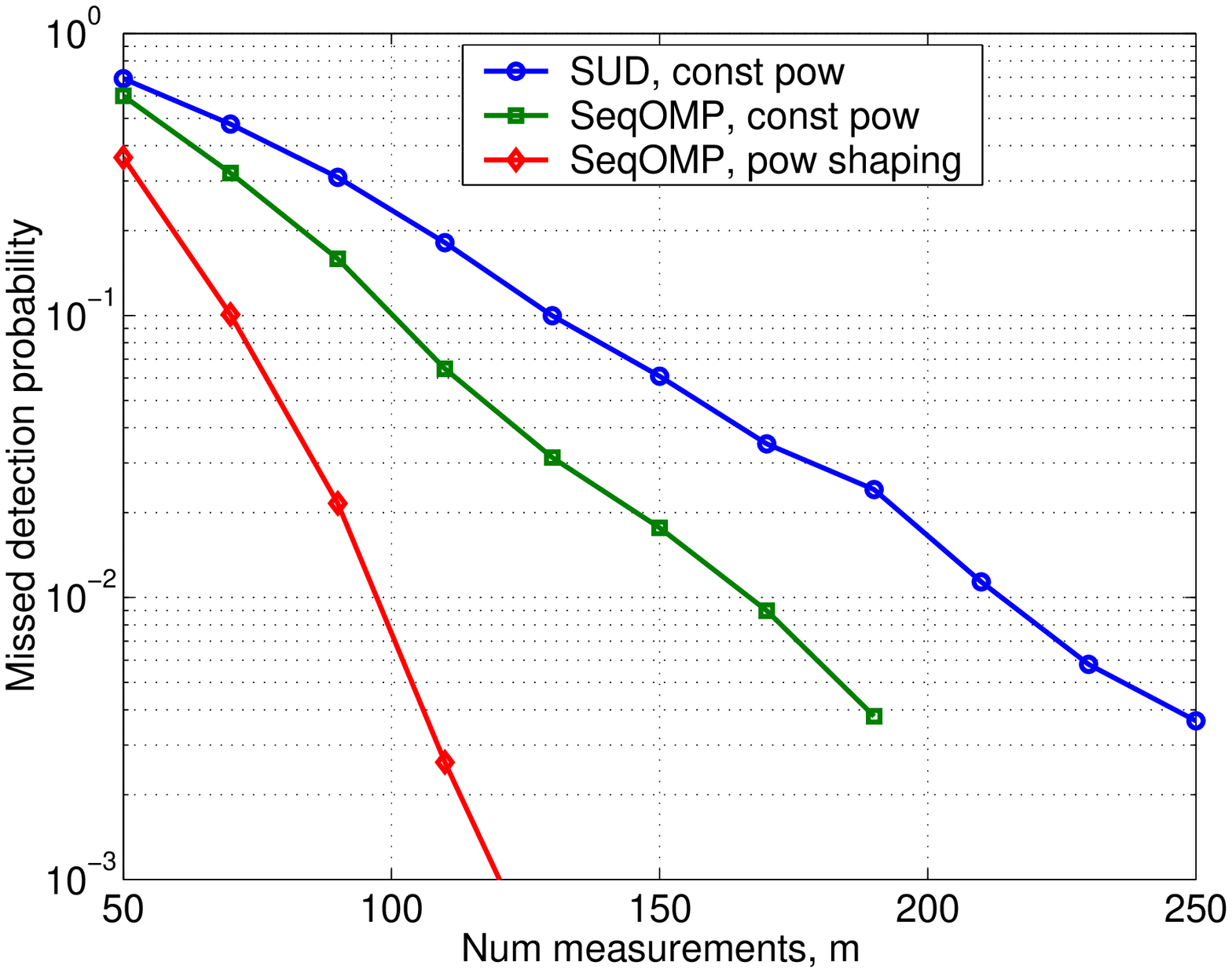,width=3.5in}
 \end{center}
 \caption{Missed detection probabilities for various
  detection methods and power profiles.
  The number of users is $n=100$, $\captionSNR = 20\,\dB$,
  the activity probability is $\lambda = 0.1$, and
  the false alarm rate is $\PFA = 10^{-3}$.
  For the SeqOMP algorithm with power shaping,
   the leakage fraction was set to $\theta = 0.1$.}
 \label{fig:sompSim}
\end{figure}

Fig.\ \ref{fig:sompSim} shows a more direct comparison of the
performance of single-user detection and SeqOMP with power shaping.
In the simulation, there are
$n =$ 100 users, the activity probability is $\lambda = 0.1$,
and the total SNR is 20 dB\@.
The number of measurements $m$ was varied, and for each $m$,
the missed detection probability was estimated with 1000 Monte Carlo
trials.

As expected, single-user detection requires the most number of measurements.
For a missed detection rate of 1\%, Fig.\ \ref{fig:sompSim} shows that
single-user detection requires approximately $m \approx 210$ measurements.
In this simulation of single-user detection,
all users arrived at the same power.
Employing SeqOMP, but keeping the power profile of the users constant,
decreases the number of measurements somewhat to $m \approx 170$ for
a 1\% missed detection rate.
However, using SeqOMP with power shaping decreases the number of measurements
by more than a factor of two to $m \approx 95$.
Thus, at least at high SNRs, SeqOMP may provide significant gains over simple
single-user detection.

\subsection{OMP with Power Shaping}

As discussed earlier, although SeqOMP can provide gains over single-user detection,
its performance is typically worse than OMP, even if SeqOMP is used with power shaping.
Our interest in the algorithm is that it is simple to analyze.  However, we can in principle
use power shaping with the better OMP algorithm as well.

\begin{figure}
 \begin{center}
  \epsfig{figure=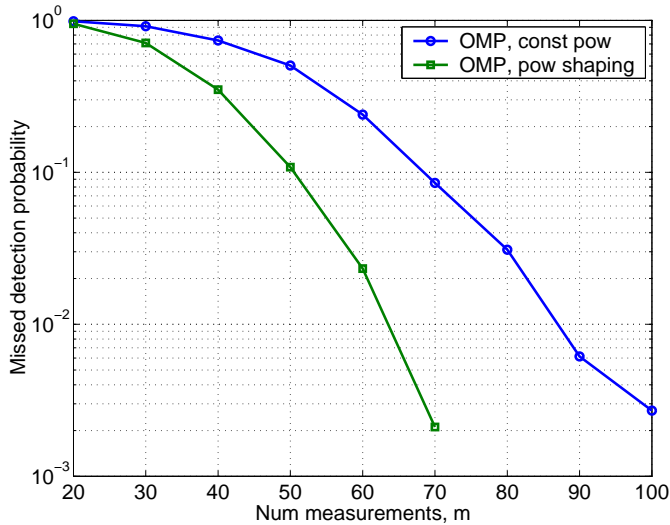,width=3.5in}
 \end{center}
 \caption{Power shaping with OMP\@.  Plotted is the missed detection
 probabilities with OMP using a constant power profile, and
 power shaping wiht a leakage fraction set to $\theta = 0.1$.
 Other simulation assumptions are identical to Fig.\ \ref{fig:sompSim}.}
 \label{fig:ompSim}
\end{figure}

While we do not have any analytical result, the simulation in Fig.~\ref{fig:ompSim}
shows that power shaping provides some gains with OMP as well.
Specifically, when the users are targeted at equal receive power,
$m \approx 85$ measurements are needed for a missed detection probability of 1\%.
This number is slightly lower than that required by SeqOMP,
even when SeqOMP uses power shaping.
When OMP is used with power shaping, the number of measurements decreases to about
$m \approx 65$.

\section{Relations to MAC Capacity} \label{sec:capacity}

As discussed in the introduction,
the random access channel is a special case of a multiple access channel (MAC).
One of the fundamental results in network information
theory \cite{Ahlswede:71,CoverT:91} is that, under certain assumptions,
the sum rate with multiple users transmitting to a single receiver
without coordination can equal the capacity with coordination.
However, one of the key assumptions in this classic result is that
the users employ capacity-achieving block codes.
In the on--off random access channel considered here,
users transmit on a single codeword and therefore
cannot benefit from channel coding.
Thus, unlike the classic MAC channel, the random access channel
may incur a loss in capacity due to the lack of coordination amongst users.

To evaluate this possibility,
let us first compute the effective ``sum rate" transmitted in the
on--off random access channel.  Each user transmits with a probability $\lambda$,
so the information conveyed in detecting the user's activity is $h(\lambda)$,
where $h(\lambda)$ is the binary entropy,
\[
    h(\lambda) = -\lambda\log(\lambda) - (1-\lambda)\log(1-\lambda) \mbox{ nats.}
\]
Since there are $n$ users, if all users can be reliably detected, the total
information rate is
\[
    R = nh(\lambda).
\]

We can compare this rate with the Shannon capacity of the channel.
If all the users coordinate their transmissions, the capacity would be identical
to a single user transmitting with the same total power.
Since the channel is AWGN with $m$ channel uses, the capacity of the channel
with a single coordinated transmission would be $C = m\log(1+\SNR)$.
If the number of measurements $m$
is selected for reliable detection, the necessary condition (\ref{eq:minMeasML})
shows that the capacity is bounded below by
\[
    C = m\log(1+\SNR) \geq \frac{\log(1+\SNR)}{\SNR}\lambda n \log((1-\lambda)n).
\]
Thus, the ratio of the sum rate to capacity is bounded above by
\[
    \frac{R}{C} \leq \frac{\log(1+\SNR)h(\lambda)}{\SNR \log((1-\lambda)n)}.
\]
This ratio represents a bound on the maximum rate without coordination
amongst the users to the maximum rate possible with coordination.
If $\lambda$ and the SNR are fixed and $n \arr \infty$,
the ratio $R/C \arr 0$.  Thus, the sum rate of the
random access channel has a fundamentally lower scaling than the
standard AWGN channel.

There is, however, one case where
the random access channel's sum rate
achieves the single-user Shannon capacity.
Suppose that the SeqOMP algorithm is used with exponential power shaping.
The sufficient condition (\ref{eq:minMeasSOMP})
shows that the number of measurements $m$ can be selected such that
the Shannon capacity is
\[
    C = m\log(1+\SNR) \approx \lambda n L(\lambda,n),
\]
where, in the approximation, we have ignored the infinitesimal $\delta$,
and the $\lambda n$ term.
In this case, the ratio of the sum rate to capacity is
\[
    \frac{R}{C} \approx \frac{h(\lambda)}{\lambda L(\lambda,n)}.
\]
Now suppose the expected number of active users is fixed to some value $k$,
and we let the activity probability scale as $\lambda(n) = k/n$.
It is easily checked that $R/C \arr 1$ as $n \arr \infty$.
Therefore, with a fixed expected number of active users, the
sum rate of the random access channel matches the Shannon capacity as the
number of user $n \arr \infty$.  Moreover, the random access capacity
can be achieved with the SeqOMP method with exponential power shaping.

In a way, this result is perhaps not surprising.
When the expected number of used is fixed to some value $k$,
and the block length $m$ scales
to infinity, the random access channel becomes identical to a standard MAC channel
with $k$ users, each transmitting on a random codebook of size $n/k$.
Moreover, the SeqOMP algorithm is precisely equivalent to the classic
SIC used in conjunction with ML detection for each user.  SIC combined
with optimal decoding for each user is known to achieve the sum rate.

The connection between the MAC channel
and sparsity detection has also been observed by Jin and Rao~\cite{JinR:08}.
Specifically, they show that OMP is clearly an analogue to the 
classic SIC method.
Moreover, they argue, at least heuristically, that if $\lambda \ll 1$,
the sum-rate $R$ achievable by OMP should approach the capacity $C$.

Our analysis of SeqOMP provides analytic evidence for these
claims by showing a specific regime where $R/C \arr 1$.
However, it also shows when this intuition fails by showing that
when the SNR and activity probability $\lambda$ are fixed,
then $R/C \arr 0$.  In this case, there is a potentially-large
gap between the MAC capacity and the sum rate in the 
random on--off channel.

\section{Conclusions}
\label{sec:concl}
Sparse signal detection is a valuable framework for understanding
multiple access on--off random signaling.
Results can provide simple capacity estimates and clarify the role of power
control and multiuser detection.
Methods such as OMP and lasso, which are widely used in sparse detection
problems, can be applied as multiuser detection methods for on--off random access channels.
Analysis shows that these methods may offer improved near--far resistance over
single-user detection in high SNRs.
Optimal ML detection may theoretically offer further gains in the high SNR regime,
but is not computationally possible.
However, some gains at high SNR may be practically achievable
through power shaping and SIC-like techniques such as OMP\@.

\appendix

\section*{Proof of Theorem \ref{thm:minMeasSOMP} }
\label{sec:proofs}

\subsection{Proof Outline}
\label{sec:proofOutline}
At a high level, the proof of Theorem~\ref{thm:minMeasSOMP} is similar
to the proof of \cite[Thm.~2]{FletcherRG:08arXiv},
the single-user detection condition (\ref{eq:minMeasSUD}).
One of the difficulties in the proof is to handle the relationships
between random events at different iterations of the SeqOMP algorithm.
To avoid this difficulty, we first show an equivalence
between the success of SeqOMP and an alternative sequence of events that
is easier to analyze.
After this simplification,
small modifications handle the cancellations of detected vectors.

Fix $n$ and define
\[
    \Itrue(j) = \left\{ ~\ell~: ~\ell \in \Itrue, \ell \leq j \right\},
\]
which is the set of elements of the active set with indices $\ell \leq j$.
Observe that $\Itrue(0) = \{\emptyset\}$ and $\Itrue(n) = \Itrue$.

Let $\Ptrue(j)$ be the projection operator onto the orthogonal complement
of $\{ \abf_\ell, ~\ell \in \Itrue(j-1)\}$, and define
\beq \label{eq:rhotrue}
    \rhotrue(j) = \frac{|\abf_j'\Ptrue(j)\ybf|^2}{\|\Ptrue(j)\abf_j\|^2\|\Ptrue(j)\ybf\|^2}.
\eeq
A simple induction argument shows that Algorithm \ref{algo:somp}
correctly detects the elements in the active set if and only if, at each
iteration $j$, the variables $\Ihat(j)$, $\Pbf(j)$ and $\rho(j)$ defined
in the algorithm are equal to $\Itrue(j)$, $\Ptrue(j)$ and $\rhotrue(j)$,
respectively.
Therefore, if we define
\beq \label{eq:IhatSOMPtrue}
    \Ihat = \left\{ ~j~: ~\rhotrue(j) > \mu ~\right\},
\eeq
then  Algorithm \ref{algo:somp} correctly detects all users if
and only if $\Ihat = \Itrue$.
In particular,
\[
    \Perr(n) = \Pr\left( \Ihat \neq \Itrue \right).
\]

To prove that $\Perr(n) \arr 0$ it suffices to show that
there exists a sequence of threshold levels $\mu(n)$
such the following two limits
\beqa
    \liminf_{n \arr \infty} \min_{j \in \Itrue(n)} \frac{\rhotrue(j)}{\mu} > 1,
        \label{eq:rhoMD} \\
    \limsup_{n \arr \infty} \max_{j \not \in \Itrue(n)} \frac{\rhotrue(j)}{\mu} < 1,
        \label{eq:rhoFA}
\eeqa
hold in probability.
The first limit (\ref{eq:rhoMD}) ensures that all the components in the
active set will not be missed and will be called the \emph{zero missed detection
condition}.
The second limit (\ref{eq:rhoFA}) ensures that all the components not
in the active set will not be falsely detected and will be called the
\emph{zero false alarm condition}.

Set the sequence of threshold levels as follows.
Since $\delta > 0$, we can find an $\epsilon > 0$ such that
\beq \label{eq:epsDef}
    (1+\delta) \geq (1+\epsilon)^2.
\eeq
For each $n$, let the threshold level be
\beq \label{eq:muDef}
    \mu = (1+\epsilon)\frac{\log(n(1-\lambda))}{m - \lambda n}.
\eeq
The asymptotic lack of missed detections and false alarms
with these thresholds are proven in Appendices~\ref{sec:pmd}
and~\ref{sec:pfa}, respectively.
In preparation for these sections,
Appendix \ref{sec:chiSq} reviews some facts concerning tail
bounds on Chi-squared and Beta random variables and Appendix
\ref{sec:prelimComp} performs some preliminary computations.

\subsection{Chi-Squared and Beta Random Variables}
\label{sec:chiSq}
The proof requires a number of simple facts concerning
chi-squared and beta random variables.
These variables are reviewed in~\cite{EvansHP:00}.
We will omit or just provide some sketches of the proofs
of the results in this section since they are all standard.

A random variable $u$ has a \emph{chi-squared} distribution with $r$
degrees of freedom if it can be written as
    $u = \sum_{i=1}^r z_i^2$,
where $z_i$ are i.i.d.\ ${\mathcal{N}}(0,1)$.
If $u$ is a chi-squared with two degrees of freedom, the
random variable $v = u/2$ has a \emph{Rayleigh distribution}.
For this work, chi-squared and Rayleigh distributed random variables
arise in two important instances.

\begin{lemma} \label{lem:chiSqDef}
Suppose $\xbf \in \C^r$ has a complex Gaussian distribution
${\mathcal{CN}}(0,\sigma^2I_r)$.  Then:
\begin{itemize}
\item[(a)] $2\|\xbf\|^2/\sigma^2$ is chi-squared with $2r$ degrees of freedom; and
\item[(b)] if $\ybf$ is any other $r$-dimensional random vector that is
nonzero with probability one and independent of $\xbf$, then the variable
\[
    u = \frac{|\xbf'\ybf|^2}{\sigma^2\|\ybf\|^2}
\]
has a Rayleigh distribution.
\end{itemize}
\end{lemma}
\begin{proof}
Part (a) follows from the fact that the norm $2\|\xbf\|^2/\sigma^2$
is a sum of squares of $2r$ unit-variance Gaussian random variables,
one for each component
of $\sqrt{2/\sigma}\,\xbf$.
Part (b) follows from the fact that $\xbf'\ybf/(\|\ybf\|\sigma)$ is a unit-variance
complex Gaussian random variable.
\end{proof}

The following two lemmas provide standard tail bounds.

\begin{lemma} \label{lem:chiSqEq}
Suppose that for each $n$, $\{\xbf_j^{(n)}\}_{j=1}^n$
is a set of complex Gaussian random vectors with each $\xbf_j^{(n)}$
spherically symmetric in an $m_j(n)$-dimensional space.
The variables may be dependent.  Suppose also that $\Exp\|\xbf_j^{(n)}\|^2 = 1$
and
\[
    \limn \log(n) / \mmin(n) = 0
\]
where
\[
    \mmin(n) = \min_{j=1,\ldots,n} m_j(n).
\]
Then the limits
\[
    \limn \max_{j=1,\ldots,n} \|\xbf_j^{(n)}\|^2
    = \limn \min_{j=1,\ldots,n}  \|\xbf_j^{(n)}\|^2 = 1
\]
hold in probability.
\end{lemma}
\begin{proof}
From Lemma \ref{lem:chiSqDef}, for every $j$ and $n$, the norms
\[
    z(j,n) = 2m_j(n)\|\xbf_j^{(n)}\|^2,
\]
are chi-squared random variables with $2m_j(n)$ degrees of freedom.
A standard tail bound (see, for example
\cite{Wainwright:07-725}), shows that for any $\epsilon > 0$,
\beqan
    \Pr\left( \frac{z(j,n)}{2m_j(n)} > 1+\epsilon \right)
    &\leq& \exp(-2\epsilon m_j(n)) \\
    &\leq& \exp(-2\epsilon \mmin(n))
\eeqan
where the last step is due to the fact that $m_j(n) \geq \mmin(n)$.
So, using the union bound,
\beqan
    \lefteqn{\Pr\left( \max_{j=1,\ldots,n}\|\xbf_j^{(n)}\|^2
                        > 1+\epsilon \right)}\\
    &=& \Pr\left( \max_{j=1,\ldots,n}\frac{z(j,n)}{2m_j(n)}> 1+\epsilon \right) \\
    &\leq& n\max_{j=1,\ldots,n}\Pr\left( \frac{z(j,n)}{2m_j(n)}> 1+\epsilon \right) \\
    &\leq& n\exp(-2\epsilon \mmin(n))  \\
      &=& \exp(-2\epsilon \mmin(n) + \log(n))\ \arr \ 0,
\eeqan
where the last step is due to the fact that $\log(n) / \mmin(n) \arr 0$.
This shows that
\[
    \limsupn \max_{j=1,\ldots,n} \|\xbf_j^{(n)}\|^2 \leq 1
\]
in probability.

Similarly, using the tail bound that
\[
    \Pr\left( \frac{z(j,n)}{2m_j(n)} < 1-\epsilon\right)
    \leq \exp(-\epsilon^2 m_j(n)),
\]
one can show that
\[
    \liminfn \min_{j=1,\ldots,n} \|\xbf_j^{(n)}\|^2 \geq 1
\]
in probability, and this proves the lemma.
\end{proof}

\begin{lemma} \label{lem:chiSqMax}  Suppose that for each $n$,
$\{u_j^{(n)}\}_{j=1}^n$ is a set of Rayleigh random variables.
The variables may be dependent.
Then
\beq \label{eq:chiSqMaxa}
    \limsupn \max_{j=1,\ldots,n} \frac{u_j^{(n)}}{\log(n)} \leq 1 ,
\eeq
where the limit is in probability.
\end{lemma}
\begin{proof}
Since each $u_j^{(n)}$ is Rayleigh, for any $\mu > 0$,
\[
    \Pr(u_j^{(n)} > \mu) = e^{-\mu}.
\]
Combining this with the union bound, we see that for any $\epsilon > 0$,
\beqan
    \lefteqn{ \Pr\left( \max_{j=1,\ldots,n} \frac{u_j^{(n)}}{\log(n)}
    > (1+\epsilon) \right) } \\
    &\leq& n\exp(-(1+\epsilon)\log(n)) = n^{-\epsilon} \arr 0.
\eeqan
This proves the limit (\ref{eq:chiSqMaxa}).
\end{proof}

The final two lemmas concern certain beta distributed random variables.
A real-valued scalar random variable $w$ follows
a $\BetaDist(r,s)$ distribution if it can be written as
    $w = u_r / (u_r + v_s)$,
where the variables $u_r$ and $v_s$ are independent chi-squared random variables
with $r$ and $s$ degrees of freedom, respectively.  The importance of the
beta distribution is given by the following lemma.

\begin{lemma} \label{lem:betaProj}
Suppose $\xbf$ and $\ybf$ are independent random $r$-dimensional complex
random vectors with $\xbf$ being spherically-symmetrically distributed
in $\C^r$ and $\ybf$ having any distribution
that is nonzero with probability one.  Then the random variable
\[
    w = \frac{|\xbf'\ybf|^2}{\|\xbf\|^2\|\ybf\|^2}
\]
is independent of $\xbf$ and follows a $\BetaDist(2,r-2)$ distribution.
\end{lemma}
\begin{proof}
This can be proven along the lines of the arguments in
\cite{FletcherRGR:06}.
\end{proof}

The following lemma provides a simple expression for the
maxima of certain beta distributed variables.

\begin{lemma} \label{lem:betaLim}
For each $n$, suppose $\{w_j^{(n)}\}_{j=1}^n$
is a set of random variables with $w_j^{(n)}$ having a
$\BetaDist(2,m_j(n)-2)$ distribution.
Suppose that
\beq \label{eq:limlogb}
    \limn \log(n) / \mmin(n) = 0, \ \ \
    \limn \mmin(n) = \infty
\eeq
where
\[
    \mmin(n) = \min_{j=1,\ldots,n} m_j(n).
\]
Then,
\[
    \limsupn \max_{j=1,\ldots,n} \frac{m_j(n)}{\log(n)}w_j^{(n)} \leq 1
\]
in probability.
\end{lemma}
\begin{proof}
We can write $w_j^{(n)} = u_j^{(n)} / (u_j^{(n)} + v_j^{(n)})$
where $u_j^{(n)}$ and $v_j^{(n)}$
are independent chi-squared random variables with 2 and $m_j(n)-2$ degrees of
freedom, respectively.
Let
\beqan
    U_n &=& \frac{1}{2\log(n)}\max_{j=1,\ldots,n} u_j^{(n)}, \\
    V_n &=& \min_{j=1,\ldots,n}\frac{1}{2m_j(n)-2} v_j^{(n)}, \\
    T_n &=& \max_{j=1,\ldots,n} \frac{m_j(n)w_j^{(n)}}{\log(n)}.
\eeqan
The condition (\ref{eq:limlogb}) and an argument similar to the proof of
Lemma~\ref{lem:chiSqEq} shows that $V_n \arr 1$ in probability.
Also, $U_n/2$ is Rayleigh distributed so Lemma~\ref{lem:chiSqMax} shows that
\[
    \limsupn U_n \leq 1
\]
in probability.
Using these two limits along with (\ref{eq:limlogb}) shows that
\beqan
    \limsupn T_n &=&\limsupn \max_{j=1,\ldots,n} \frac{m_j(n)w_j^{(n)}}{\log(n)} \\
    &=&\limsupn \max_{j=1,\ldots,n} \frac{m_j(n)}{\log(n)}
    \frac{u_j^{(n)}}{u_j^{(n)} + v_j^{(n)}} \\
    &=& \limsupn \frac{U_n}{V_n + U_n\log(n)/m_j(n)} \\
         &\leq& \frac{1}{1 + (1)(0)} = 1,
\eeqan
where the limit is in probability.
\end{proof}

\subsection{Preliminary Computations and Technical Lemmas}
\label{sec:prelimComp}
We first need to prove a number of simple but technical bounds.
We begin by considering the dimension $m_i$ defined as
\beq \label{eq:miDef}
    m_i = \dim(\range(\Ptrue(i))).
\eeq
Our first lemma computes the limit of this dimension.

\begin{lemma} \label{lem:miMin}
The following limit
\beq \label{eq:miMin}
    \lim_{n \arr \infty} \min_{i=1,\ldots,n} \frac{m_i}{m - \lambda n} = 1
\eeq
holds in probability and almost surely.  The deterministic limits
\beq \label{eq:lognm}
    \lim_{n \arr \infty} \frac{\log(\lambda n)}{m - \lambda n}
    = \lim_{n \arr \infty} \frac{\log((1-\lambda) n)}{m - \lambda n} = 0
\eeq
also hold.
\end{lemma}
\begin{proof}
Recall that $\Ptrue(i)$ is the projection onto the orthogonal complement
of the vectors $\abf_j$ with $j \in \Itrue(i-1)$.
With probability one, these vectors will be linearly independent, so $\Ptrue(i)$
will have dimension $m - |\Itrue(i-1)|$.  Since $\Itrue(i)$ is increasing with $i$,
\beqa
    \min_{i=1,\ldots,n} m_i &=& m - \max_{i=1,\ldots,n} |\Itrue(i-1)| \nonumber \\
    &=& m - |\Itrue(n-1)|. \label{eq:miItrue}
\eeqa
Since each user is active with probability $\lambda$ and the activities of the
users are independent, the law of large numbers shows that
\[
    \lim_{n \arr \infty} \frac{|\Itrue(n-1)|}{\lambda(n-1)} = 1
\]
in probability and almost surely.  Combining this with (\ref{eq:miItrue})
shows (\ref{eq:miMin}).

We next show (\ref{eq:lognm}).
Since the hypothesis of the theorem requires that
$\lambda n$, $(1-\lambda)n$ and $m -\lambda n$ all approach infinity,
the fractions in (\ref{eq:lognm}) are eventually positive.
Also, from (\ref{eq:Ldef}), $L(\lambda,n) < \max\{\log(\lambda n), \log((1-\lambda)n)\}$.
Therefore, from (\ref{eq:minMeasGam}),
\beqan
    \lefteqn{ \frac{1}{m-\lambda n}\max\{\log(\lambda n), \log((1-\lambda)n)\} }\\
    &\leq& \frac{\gamma}{L(\lambda,n)}\max\{\log(\lambda n), \log((1-\lambda)n)\}
    \leq \gamma \arr 0,
\eeqan
where the last step is from the hypothesis of the theorem.
\end{proof}

\medskip
Next, for each $i=1,\ldots,n$, define the \emph{residual vector},
\beq \label{eq:residDef}
    \ebf_i = \Ptrue(i)(\ybf - \abf_ix_i).
\eeq
Observe that
\beqa
    \ebf_i &=& \Ptrue(i)(\ybf - \abf_ix_i) \nonumber \\
        &\stackrel{(a)}{=}& \Ptrue(i)\left(\wbf + \sum_{j \neq i} \abf_jx_j \right)
        \nonumber \\
        &\stackrel{(b)}{=}& \Ptrue(i)\left(\wbf + \sum_{j > i} \abf_jx_j \right)
        \label{eq:residSum}
\eeqa
where (a) follows from (\ref{eq:yax}) and
(b) follows from the fact that $\Ptrue(i)$ is the projection onto
the orthogonal complement of the span of all vectors $\abf_j$ with $j < i$
and $x_j \neq 0$.

The next lemma shows that the power of the residual
vector is described by the random variable
\beq\label{eq:sigDef}
    \sigma^2(i) = 1 + \sum_{j = i+1}^n |x_j|^2.
\eeq

\medskip
\begin{lemma} \label{lem:residDist}
For all $i=1,\ldots,n$, the residual vector $\ebf_i$,
conditioned on the modulation vector $\xbf$ and projection $\Ptrue(i)$,
is a spherically symmetric Gaussian in the range space of $\Ptrue(i)$
with total variance
\beq \label{eq:residPow}
    \Exp\left(\|\ebf_i\|^2 \mid \xbf \right) = \frac{m_i}{m}\sigma^2(i),
\eeq
where $m_i$ and $\sigma^2(i)$ are defined in (\ref{eq:miDef}) and (\ref{eq:sigDef}),
respectively.
\end{lemma}
\begin{proof}
Let
\[
    \vbf_i = \wbf + \sum_{j > i} \abf_jx_j,
\]
so that $\ebf_i = \Ptrue(i)\vbf_i$.
Since the vectors $\abf_j$ and $\wbf$ have Gaussian
${\mathcal{CN}}(0,1/mI_m)$ distributions, for a given modulation vector $\xbf$,
$\vbf_i$ must be a zero-mean white Gaussian vector with total variance
$\Exp\|\vbf_i\|^2 = \sigma^2(i)$.
Also, since the operator $\Ptrue(i)$ is a function of the components
$x_\ell$ and vectors $\abf_\ell$ for $\ell < i$,
$\Ptrue(i)$ is independent of the vectors $\wbf$ and $\abf_j$, $j > i$,
and therefore independent of $\vbf_i$.
Since $\Ptrue(i)$ is a projection from an $m$-dimensional space
to an $m_i$-dimensional space, $\ebf_i$, conditioned on the modulation
vector $\xbf$, must be spherically symmetric Gaussian in the range space
of $\Ptrue(i)$ with total variance satisfying (\ref{eq:residPow}).
\end{proof}

Our next lemma requires the following version of
the well-known Hoeffding's inequality.

\begin{lemma}[Hoeffding's Inequality] \label{lem:hoeffding}
Suppose $z$ is the sum
\[
    z = z_0 + \sum_{i=1}^r z_i
\]
where $z_0$ is a constant and the variables $z_i$ are independent random variables
that are almost surely bounded in some interval $z_i \in [a_i,b_i]$.
Then, for all $\epsilon > 0$,
\[
    \Pr\left( z - \Exp(z) \geq \epsilon \right) \leq
        \exp\left(\frac{-2\epsilon^2}{C}\right),
\]
where
\[
    C = \sum_{i=1}^r (b_i-a_i)^2.
\]
\end{lemma}
\begin{proof}  See \cite{Hoeffding:63}.
\end{proof}

\begin{lemma} \label{lem:sigLim}
Under the assumptions of Theorem \ref{thm:minMeasSOMP}, the limit
\[
    \limsup_{n \arr \infty} \max_{i=1,\ldots,n} \frac{\sigma^2(i)}{\sigmahat^2(i)} \leq 1
\]
holds in probability.
\end{lemma}
\begin{proof}
Let $z(i) = \sigma^2(i)/\sigmahat^2(i)$.
From the definition of $\sigma^2(i)$ in (\ref{eq:sigDef}), we can write
\[
    z(i) = \frac{1}{\sigmahat^2(i)} + \sum_{j=i+1}^n z(i,j),
\]
where $z(i,j) = |x_j|^2/\sigmahat^2(i)$ for $j > i$.

Now recall that in the problem formulation,
each user is active with probability $\lambda$,
with power $|x_j|^2 = p_j$ conditioned on when the user being active.
Also, the activities of different users are independent,
and the conditional powers $p_j$ are treated as deterministic
quantities.
Therefore, the variables $z(i,j)$ are independent with
\[
    z(i,j) = \left\{ \begin{array}{ll}
        p_j/\sigmahat^2(i), & \mbox{with probability $\lambda$;} \\
        0,                  & \mbox{with probability $1-\lambda$},
        \end{array} \right.
\]
for $j > i$.
Combining this with the definition of $\sigmahat^2(i)$ in (\ref{eq:sigHatDef}),
we see that
\[
    \Exp(z(i)) = \frac{1}{\sigmahat^2(i)}\left(1 + \lambda \sum_{j=i+1}^n p_j\right) = 1.
\]
Also, for each $j > i$, we have the bound
\[
    z(i,j) \in [0, p_j/\sigmahat^2(i)].
\]
So for use in Hoeffding's Inequality (Lemma~\ref{lem:hoeffding}),
define
\[
    C = C(i,n) = \sigmahat^{-4}(i)\sum_{j = i+1}^n p_j^2,
\]
where dependence of the power profile and $\sigmahat(i)$ on $n$ is implicit.
Now define
\[
    c_n = \max_{i=1,\ldots,n} \log(n) C(i,n),
\]
so that $C(i,n) \leq c_n/\log(n)$ for all $i$.
Hoeffding's Inequality (Lemma~\ref{lem:hoeffding}) now shows
that for all $i < n$,
\beqan
    \Pr(z(i) \geq 1+\epsilon) &\leq& \exp\left(-2\epsilon^2/C(i,n)\right)  \\
        &\leq& \exp\left(-2\epsilon^2\log(n)/c_n\right).
\eeqan
Using the union bound,
\beqan
    \lefteqn{\limn \Pr\left( \max_{j=1,\ldots,n} z(i) > 1 + \epsilon\right) } \\
    &\leq&
       \limn  n\exp\left(- \frac{2\epsilon^2\log(n)}{c_n}\right) \\
     &=&  \limn  n^{1- 2\epsilon^2/c_n} = 0.
\eeqan
The final step is due to the fact that the technical condition
(\ref{eq:pjsqbnd}) in the theorem implies $c_n \arr 0$.  This proves the lemma.
\end{proof}

\subsection{Missed Detection Probability}
\label{sec:pmd}

Consider any $j \in \Itrue$.
Using (\ref{eq:residDef}) to rewrite (\ref{eq:rhotrue})
along with some algebra shows
\beqa
    \rhotrue(j) &=& \frac{|\abf_j'\Ptrue(j)\ybf|^2}{\|\Ptrue(j)\abf\|^2\|\Ptrue(j)\ybf_j\|^2}
        \nonumber \\
        &=& \frac{|\abf_j'(x_j\Ptrue(j)\abf_j + \ebf_j)|^2}
        {\|\Ptrue(j)\abf_j\|^2\|x_j\Ptrue(j)\abf_j+\ebf_j\|^2} \nonumber \\
           &\geq& \frac{s_j - 2\sqrt{z_js_j} + z_j}{s_j + 2\sqrt{z_js_j} + 1},
           \label{eq:rhotrueMD}
\eeqa
where
\beqa
    s_j &=& \frac{|x_j|^2\|\Ptrue(j)\abf_j\|^2}{\|\ebf_j\|^2}, \label{eq:sjdef} \\
    z_j &=& \frac{|\abf_j'\Ptrue(j)\ebf_j|^2}{\|\Ptrue(j)\abf_j\|^2\|\ebf_j\|^2}.
         \label{eq:zjdef}
\eeqa
Define
\[
    \smin = \min_{j \in \Itrue} s_j, \ \ \
    \zmax = \max_{j \in \Itrue} z_j.
\]
We will now bound $\smin$ from below and $\zmax$ from above.

We first start with $\smin$.   Conditional on $\xbf$ and $\Ptrue(j)$,
Lemma \ref{lem:residDist} shows that each $\ebf_j$
is a spherically-symmetrically distributed Gaussian on the $m_j$-dimensional
range space of $\Ptrue(j)$.  Since there are asymptotically $\lambda n$ elements
in $\Itrue$, Lemma \ref{lem:chiSqEq} along with (\ref{eq:lognm}) show that
\beq \label{eq:sjlima}
    \lim_{n \arr \infty} \max_{j \in \Itrue} \frac{m}{m_j\sigma^2(j)}\|\ebf_j\|^2 = 1,
\eeq
where the limit is in probability.
Similarly, $\Ptrue(j)\abf_j$ is also a spherically-symmetrically distributed Gaussian
in the range space of $\Ptrue(j)$.  Since $\Ptrue(j)$ is a projection from an $m$-dimensional
space to a $m_j$-dimensional space and $\Exp\|\abf_j\|^2 = 1$, we have that
$\Exp\|\Ptrue(j)\abf_j\|^2 = m_j/m$.
Therefore, Lemma~\ref{lem:chiSqEq} along with (\ref{eq:lognm}) show that
\beq \label{eq:sjlimb}
    \lim_{n \arr \infty} \min_{j \in \Itrue} \frac{m}{m_j}\|\Ptrue(j)\ebf_j\|^2 = 1.
\eeq
Taking the limit (in probability) of $\smin$,
\beqa
    \liminfn \frac{\smin}{\gamma} &=& \liminfn \min_{j \in \Itrue} \frac{s_j}{\gamma}
        \nonumber \\
    &\stackrel{(a)}{=}& \liminfn \min_{j \in \Itrue}
        \frac{|x_j|^2\|\Ptrue(j)\abf_j\|^2}{\gamma\|\ebf_j\|^2} \nonumber \\
    &\stackrel{(b)}{=}& \liminfn \min_{j \in \Itrue}
        \frac{|x_j|^2}{\gamma\sigma^2(j)} \nonumber \\
    &\stackrel{(c)}{=}& \liminfn \min_{j \in \Itrue}
        \frac{p_j}{\gamma\sigma^2(j)} \nonumber \\
    &\stackrel{(d)}{\geq}& \liminfn \min_{j \in \Itrue}
        \frac{p_j}{\gamma\sigmahat^2(j)} \nonumber \\
         &\stackrel{(e)}{\geq}& 1, \label{eq:sminLim}
\eeqa
where (a) follows from (\ref{eq:sjdef});
(b) follows from (\ref{eq:sjlima}) and (\ref{eq:sjlimb});
(c) follows from (\ref{eq:pjdef});
(d) follows from Lemma \ref{lem:sigLim};
and (e) follows from (\ref{eq:sinrDef}).

We next consider $\zmax$.  Conditional on $\Ptrue(j)$, the vectors $\Ptrue(j)\abf_j$ and
$\ebf_j$ are independent spherically-symmetric complex Gaussians in the
range space of $\Ptrue(j)$.  It follows from Lemma~\ref{lem:betaProj} that
each $z_j$ is a $\BetaDist(2,m_j-2)$ random variable.
Since there are asymptotically
$\lambda n$ elements in $\Itrue$, Lemma \ref{lem:betaLim} along
with (\ref{eq:miMin}) and (\ref{eq:lognm}) show that
\beq
    {\limsupn \frac{m - \lambda n}{\log(\lambda n)} \zmax}
    =    \limsupn \frac{m - \lambda n}{\log(\lambda n)} \max_{j \in \Itrue}
        z_j \leq 1. \label{eq:zmaxLim}
\eeq

The above analysis shows that for any $j \in \Itrue$,
\beqa
    \lefteqn{ \liminfn \min_{j \in \Itrue} \frac{1}{\sqrt{\mu}}(\sqrt{s_j} - \sqrt{z_j}) } \nonumber \\
    &\stackrel{(a)}{\geq}& \liminfn \frac{1}{\sqrt{\mu}}(\sqrt{\smin} - \sqrt{\zmax}) \nonumber \\
    &\stackrel{(b)}{\geq}& \liminfn \frac{1}{\sqrt{\mu}} \left(\sqrt{\gamma}
       - \sqrt{\frac{\log(\lambda n)}{m-\lambda n}} \right) \nonumber  \\
    &\geq& \liminfn \sqrt{\frac{1+\delta}{\mu}} \left(\sqrt{\frac{\gamma}{1+\delta}}
       - \sqrt{\frac{\log(\lambda n)}{m-\lambda n}} \right) \nonumber  \\
    &\stackrel{(c)}{\geq}& \liminfn \sqrt{\frac{1+\delta}{(m-\lambda n)\mu}}
        \left(\sqrt{L(\lambda,n)}
       - \sqrt{\log(\lambda n)} \right) \nonumber  \\
    &\stackrel{(d)}{=}& \liminfn \sqrt{\frac{(1+\delta)\log(n(1-\lambda))}{(m-\lambda n)\mu}}
        \nonumber  \\
    &\stackrel{(e)}{=}& \liminfn \sqrt{\frac{1+\delta}{1+\epsilon}} \nonumber \\
    &\stackrel{(f)}{\geq}& \sqrt{1+\epsilon} \label{eq:szdiff}
\eeqa
where (a) follows from the definitions of $\smin$ and $\zmax$;
(b) follows from (\ref{eq:sminLim}) and (\ref{eq:zmaxLim});
(c) follows from (\ref{eq:minMeasGam});
(d) follows from (\ref{eq:Ldef});
(e) follows from (\ref{eq:muDef}); and
(f) follows from (\ref{eq:epsDef}).
Therefore, starting with (\ref{eq:rhotrueMD}),
\beqan
    \lefteqn{
    \liminf_{n \arr \infty} \min_{j \in \Itrue}
    \frac{\rho(j)}{\mu} } \nonumber \\
    & \stackrel{(a)}{\geq}&
    \liminfn  \min_{j \in \Itrue} \frac{1}{\mu}
        \frac{s_j - 2\sqrt{z_js_j} + z_j}{s_j + 2\sqrt{z_js_j} + 1} \nonumber \\
    & =&
    \liminfn  \min_{j \in \Itrue} \frac{1}{\mu}
        \frac{(\sqrt{s_j} - \sqrt{z_j})^2}{s_j + 2\sqrt{z_js_j} + 1} \nonumber \\
    & \stackrel{(b)}{\geq}&
    \liminfn  \min_{j \in \Itrue}
        \frac{1+\epsilon}{s_j + 2\sqrt{z_js_j} + 1} \nonumber \\
    & \stackrel{(c)}{\geq}&
    \liminfn  \min_{j \in \Itrue}
        \frac{1+\epsilon}{s_j + 2\sqrt{s_j} + 1} \nonumber \\
    & \geq&
    \liminfn  \min_{j \in \Itrue}
        \frac{1+\epsilon}{\smin + 2\sqrt{\smin} + 1} \nonumber \\
    & \stackrel{(d)}{\geq}&
     \liminfn  \min_{j \in \Itrue}
        \frac{1+\epsilon}{(\sqrt{\gamma} + 1)^2}  \stackrel{(e)}{=}1+\epsilon,
\eeqan
where (a) follows from (\ref{eq:rhotrueMD});
(b) follows from (\ref{eq:szdiff});
(c) follows from the fact that $z_j \in [0,1]$ (it is a Beta distributed random variable);
(d) follows from (\ref{eq:sminLim}); and
(e) follows from the condition of the hypothesis of the theorem that $\gamma \arr 0$.
This proves the first requirement, condition (\ref{eq:rhoMD}).

\subsection{False Alarm Probability}
\label{sec:pfa}

Now consider any index $j \not \in \Itrue$.  This implies that $x_j=0$ and therefore
(\ref{eq:residDef}) shows that
\[
    \Ptrue(j)\ybf
     = \ebf_j.
\]
Hence from (\ref{eq:rhotrue}),
\beq \label{eq:rhozj}
   \rhotrue(j) = \frac{|\abf_j'\ebf|^2}
                      {\|\Ptrue(j)\abf\|^2\|\ebf_j\|^2}
   = z_j
\eeq
where $z_j$ is defined in (\ref{eq:zjdef}).  From the discussion above,
each $z_j$ has the $\BetaDist(2,m_j-2)$ distribution.
Since there are asymptotically $(1-\lambda) n$ elements in $\Itrue^c$,
the conditions (\ref{eq:miMin}) and (\ref{eq:lognm}) along with Lemma \ref{lem:betaLim}
show that the limit
\beq \label{eq:zbndfa}
    \limsupn \max_{j \not\in \Itrue} \frac{m-\lambda n}{\log(n(1-\lambda n)}z_j \leq 1
\eeq
holds in probability.
Therefore,
\beqan
    \lefteqn{ \limsupn \max_{j \not\in \Itrue} \frac{1}{\mu}\rhotrue(j) } \\
    &\stackrel{(a)}{=}& \limsupn \max_{j \not\in \Itrue} \frac{1}{\mu}z_j  \\
    &\stackrel{(b)}{=}& \limsupn \max_{j \not\in \Itrue} \frac{m - \lambda n}{(1+\epsilon)
            \log(n(1-\lambda n)}z_j \\
    &\stackrel{(c)}{\leq}& \frac{1}{1+\epsilon}
\eeqan
where (a) follows from (\ref{eq:rhozj});
(b) follows from (\ref{eq:muDef}); and
(c) follows from (\ref{eq:zbndfa}).
This proves (\ref{eq:rhoFA}) and thus completes the proof of the theorem.

\section*{Acknowledgments}
The authors thank Martin Vetterli for his support, wisdom, 
and encouragement.
The authors also thank Gerhard Kramer for helpful comments on an early draft
of this manuscript.

\bibliographystyle{IEEEtran}
\bibliography{bibl}

\end{document}